# Picosecond Timing Resolution Detection of Gamma Photons Utilizing Microchannel-plate Detectors: Experimental Tests of Quantum Nonlocality and Photon Localization

Victor D. Irby

*Department of Physics, University of South Alabama, Mobile, Al 36688-0002*

(Dated: March 18, 2004)

## Abstract

The concept and subsequent experimental verification of the proportionality between pulse amplitude and detector transit time for microchannel plate detectors is presented. This discovery has led to considerable improvement in the overall timing resolution for detection of high energy gamma photons. Utilizing a $^{22}$Na positron source, a full width half maximum (FWHM) timing resolution of 138 ps has been achieved. This FWHM includes detector transit-time spread for *both* chevron-stack type detectors, timing spread due to uncertainties in annihilation location, all electronic uncertainty, and any remaining quantum mechanical uncertainty. The first measurement of the *minimum quantum uncertainty* in the time interval between detection of the two annihilation photons is reported. The experimental results give strong evidence *against* instantaneous spatial-localization of $\gamma$-photons due to measurement-induced nonlocal quantum wave-function collapse. The experimental results are also the first that imply momentum is conserved only after the quantum uncertainty in time has elapsed [H. Yukawa, *Proc. Phys. -Math. Soc. Japan*, **17**, 48 (1935)].

## I. INTRODUCTION

In this paper, we present high-resolution measurements of the time interval between detection of the two photons emitted in positron/electron annihilation. Although there exists extensive measurements involving positron lifetime studies in the literature, the existing studies involve measurements of the time interval between the prompt decay photon (signalling the emission of a positron) and the detection of *one* of the subsequent annihilation photons. Measurements presented here, however, focus on the time interval "*between the two annihilation photons*", which provides an important experimental test of our understanding of the annihilation process itself and our understanding of fundamental quantum mechanics. The fundamental question addressed in this paper is "*What is the minimum quantum uncertainty in the time interval between detection of the two annihilation photons?*".

We first would like to emphasize that the time uncertainty in question is **not** simply due to uncertainty in annihilation location. Let us assume that an annihilation event takes place between two opposing detectors ($D_1$ and $D_2$) facing each other with a separation equal to $L$. If the annihilation event occurs at a distance $s$, measured relative to $D_1$, the time interval $\tau$ between detection of the two photons is then given by $\tau = (t_2 - t_1) = (L - 2s)/c$. Obviously, uncertainty in annihilation location $\Delta s$ will lead to uncertainty in the measured time interval $\Delta \tau = -2\Delta s/c$. However, positrons emitted from $^{22}$Na are completely attenuated by $\approx 1mm$ of aluminum. Thus, one can easily "spatially-restrict" all annihilation events by sandwiching the source between two 1 mm aluminum plates. If a micro-gram source is painted on one of the plates, this will yield a maximum uncertainty in arrival time $\Delta \tau \approx 10ps$ which is a full order of magnitude less than the minimum lifetime of positronium in vacuum. Therefore, uncertainties due to annihilation location can be easily minimized and **are fully accounted for**.

In regard to the fundamental question presented above, let us assume that all annihilation events take place exactly halfway between $D_1$ and $D_2$. One may initially assume that the minimum uncertainty in time, in the interval between detection of the two photons, should be approximately equal to the lifetime of the positron. However, detection of one of the annihilation photons, used as a "start" in the timing measurement, is associated with the *annihilation event itself* and not with the initial creation of the positron.

The most widely accepted viewpoint is that the minimum quantum uncertainty in time,

between the two photons, **should be zero**. This assumption is based on detection-induced nonlocal collapse of the photon spatial-wavefunction, as first predicted by Einstein, Podolsky, and Rosen in their now famous "EPR Paradox" paper [1]. According to EPR, measurement of particle one's momentum enables one to infer the momentum of particle two, without interacting with particle two. Likewise, measurement of particle one's position enables one to infer the position of particle two. Thus, by measuring the position of particle one while simultaneously measuring the momentum of particle two, one then obtains knowledge of *both* particle two's momentum and position with certainty, in violation of the Heisenberg Uncertainty Principle. One is then left with the choice that either quantum mechanics is incomplete (hidden variables), or that nonlocal "spooky action at a distance" occurs. Hence the paradox. (A "bi-photon" picture has been proposed by Rubin *et. al.* [2] where no "spooky action at a distance" is required to explain the results).

An experimental test to determine the validity of "hidden variables", was first proposed in 1964 by J. S. Bell [3]. Subsequent experimental measurements, based on tests of Bell's inequalities, have strongly ruled out the existence of hidden variables and indicate that nature is fundamentally *nonlocal*. That is, a measurement of any attribute of one particle will instantaneously affect corresponding attributes of another distant particle, regardless of how far the two particles are separated. Quantum nonlocality in measurements involving polarization and two-photon interference of correlated photons has been confirmed at many independent laboratories [4–7]. In fact, the nonlocal nature of the subatomic world has recently led to new technologies such as Quantum Cryptography, Quantum Computing, and Quantum Teleportation [8]. (However, none of the experimental tests to date are completely free from "loopholes").

It has thus been generally postulated that nonlocal effects may also occur in regard to the spatial wavefunctions of emitted photons. As an example, detection of one of the photons produced in parametric down conversion is predicted to cause "instantaneous" localization of the other photon, subsequently eliminating any uncertainty in the time of arrival of the second photon at a second detector. Experimental support of this prediction has been reported by Hong *et al.* [7]. The two-photon interference method utilized in Ref. [7] indicates that the minimum time uncertainty, in the time interval between detection of the two down-converted photons, is less than $100fs$. This uncertainty in time is *much less* than the coherence time of the initial pump photons, which subsequently gives strong indication of nonlocal collapse of the photon wave-function.

One may therefore expect to observe similar nonlocal effects involving photons emitted from positron-electron annihilation. Utilizing new improvements in time resolution achieved in this laboratory, the coherence time of annihilation photons is now within the resolving power of microchannel-plate detectors. The final experimental results (See Section **V**) indicate that the *absolute minimum* uncertainty in detection time between the arrival of the two photons is not zero, but rather, $123 \pm 22\,ps$, which, quite surprisingly, agrees with the lifetime of singlet-state positronium atoms in vacuum ($125\,ps$). The exact connection with positronium lifetime, however, is not fully understood at this time.

Although nonlocal effects are observed to occur in the case of down-converted photons, the experimental results presented here give strong evidence *against* the instantaneous spatial-localization of high energy $\gamma$-photons. Despite this difference, the results presented here are found to be consistent with *both* conservation of energy and the Heisenberg Uncertainty Principle. To illustrate this point, let us assume that nonlocal instantaneous spatial-localization *does* occur for $\gamma$-photons. By detecting one photon, the uncertainty in arrival time of the second (other) photon would then be zero due to its instantaneous localization. Since the second photon is thus *well localized*, this immediately implies that the second photon would also have an *infinite uncertainty in energy*. Thus, if one measured the second photon's energy, they could obtain a value of 1 eV, or 1000 GeV, with the same probability as 0.511 MeV, which **violates conservation of energy**.

The $\gamma$-photon results presented here are found to be in agreement with recent quantum-mechanical calculations of Irby [9], which involve inclusion of *time dependence* and *conservation of energy* into the original EPR calculations.

(Note: in Ref. [9], a minimum experimental quantum uncertainty of $117 \pm 9\,ps$ was reported. This result has been updated, following additional measurement and analysis, and is reported in this paper).

## II.     OVERVIEW OF EXPERIMENTAL APPARATUS

A simplified schematic of the detection and timing circuitry is illustrated in Figure 1. The microchannel plate detectors (MCP's) consist of two identical chevron stacks ($10\,\mu$m pore diameter and L/D = 40) each with an active circular area of $4.9\,\text{cm}^2$ and each equipped with metallic anodes. (The anodes are flat circular metallic plates and are not geometrically impedance matched). The detectors are mounted in a vacuum chamber, with pressures in the low $10^{-7}$ Torr range, facing each other with a separation of $L = 10\,\text{cm}$.

The $^{22}$Na source is a standard $10\,\mu$Ci laboratory source encapsulated within a plastic disk. The source is positioned between the MCP detectors located at a distance "$s$" measured from the "start" detector. Obviously, uncertainty in annihilation location "$s$" will lead to uncertainty in the detection time interval between the photon arrival times ($\Delta t_s = 2\Delta s/c$). However, $\Delta s$ can be easily restricted by attaching sheets of aluminum to both sides of the source. (The encapsulated source has a thickness of 5 mm). Since positrons are completely attenuated by 1 mm of aluminum, all annihilation events are restricted within a 7 mm range in "$s$", resulting in a "maximum" expected uncertainty $\Delta t_s = 47$ps. (In order to prevent *any* possibility of positrons exiting from the source and striking the detectors, up to 3 mm of aluminum is actually attached to both sides of the source).

Electrical signals from the MCP anodes are passed through SHV vacuum feedthroughs into pulse pick-off circuits located outside of the vacuum vessel. (These circuits are located outside the vacuum chamber for easy access). The pick-off circuits, illustrated in more detail in Figure 2, provide pulse shaping and impedance matching before being passed into the 9327 amplifiers. The 9327 amplifiers (manufactured by PerkinElmer Ortec) **are specifically designed for MCP detectors**. The amplifiers provide both amplification/discrimination and constant-fraction timing in one unit. The 9327's convert the MCP anode signals into fast NIM output pulses for electronic timing.

The fast NIM timing signals are used as start and stop signals for the 9308 picosecond timing analyzer (PTA) (also manufactured by PerkinElmer Ortec). The 9308 time analyzer has a built in time-to-amplitude (TAC) converter, microprocessor, and internal memory. It connects to a personal computer to form a high resolution time spectrometer, capable of providing a real-time histogram of start/stop events with 1.221 ps resolution while operating in the "histogram" mode. (Because typical single detector count rates are on the order of 75 s$^{-1}$, dead time corrections are not necessary).

One major influence on timing resolution in microchannel-plate detectors is the variation in transit time, or transit time spread (TTS), of charge as it travels and multiplies through the microchannel plate (MCP) detector. The charge multiplication and transit time depends upon the energy and angle of the initiating charged particle or photon, the angular and energy distribution of secondary electrons, and the number of subsequent cascading collisions within the MCP channel. Ito, Kume, and Oba [10] studied Monte Carlo computer simulations of this process in 1984. The simulations of Ito *et al.* were carried out for a single-stage MCP with a channel diameter of D=12-$\mu$m and plate thickness of L=480-$\mu$m (L/D=40). Their analysis showed that the average transit time was 190-ps, with a FWHM spread (TTS) of 60-ps. The simulation results are in fair agreement with subsequent experimental measurements of Young *et al.* [11]. The experiments utilized an MCP with D=12-$\mu$m and L/D=40 and resulted in an average transit time of 150-ps with a TTS of 70-ps. The measurements of Young *et al.* involved detection of photons emitted from a 0.25 $\mu$m pulsed laser while gating the MCP with a 180-ps high voltage pulse (-800 V). However, in normal operation, the high voltage across the MCP is usually held constant. Thus, one may expect that the TTS measured by Young *et al.* may be somewhat overestimated. This observation is supported by experimental measurements in 1988 by Kume *et al.* [12], which involved a constant MCP bias voltage. The experiments of Kume *et al.* utilized the beam from a cavity dumped pulsed dye laser (6-ps pulse) which was subsequently split by an optical mirror. The fundamental beam, detected by a fast photodiode, was used as the start pulse, while the stop pulse was generated by the secondary beam impinging on the MCP. The resultant TTS measurements obtained by Kume *et al.* were 42-ps and 28-ps for MCP's with channel diameters of 12-$\mu$m and 6-$\mu$m, respectively.

We have recently discovered a method, utilizing existing electronic techniques, that enables one to "electronically select" a range in detector-transit times, which dramatically improves the timing resolution in microchannel-plate detectors. Currently, a full width half maximum (FWHM) timing resolution of 138 ps has been achieved. This FWHM includes TTS for *both* chevron-stack type detectors, uncertainty due to the annihilation location $\Delta s$, all electronic uncertainty, and any remaining quantum mechanical uncertainty. A discussion of the new methods is given in Section **IV**.

III.   ESTIMATION OF COINCIDENCE RATE

The $^{22}$Na isotope (half-life 2.6 years) decays by positron emission and electron capture into the 1.274 MeV excited nuclear state of $^{22}$Ne. Ninety percent of these decay events occur through positron emission. Thus, in essentially all decay events, a 1.274 MeV prompt or "decay" photon is emitted, with a 90 % chance of being followed by the emission of a positron.

Since the detectors are unable to distinguish photon energies, three types of coincident events may be detected. Because the detectors are situated 180° from each other, any coincidence between annihilation photons will essentially only involve 0.511 MeV photons. (Due to detector geometry, the probability of obtaining a coincidence between two of the three annihilation photons emitted from the triplet channel is negligible compared to that of the singlet channel). However, a prompt 1.274 MeV decay photon can also initiate a start signal, followed by an annihilation-photon stop signal. In this case, the annihilation photon may have an energy of 0.511 MeV or may have any energy ranging from zero up to 0.511 MeV, depending on whether the photon is emitted from the singlet or triplet annihilation channel. (The triplet channel emits 3 photons in a continuum of energies ranging from zero to 0.511 MeV). Thus, our estimate of overall coincidence rate must include annihilation/annihilation (AA) coincidence, decay/singlet-annihilation coincidence ($DA_1$), and decay/triplet-annihilation ($DA_3$) coincidence. However, as we shall show, the overall coincidence rate is dominated by annihilation/annihilation events when the source is situated at midpoint between the detectors.

Absolute detector efficiencies can be obtained from total single detector count rates. The total count rate for a single detector is the sum of count rates for decay photons and annihilation photons from both singlet and triplet annihilation channels. Detection efficiencies for high energy x-ray photons have been reported by the manufacturer of the micro-channel plates (Burle Industries). The MCP detection efficiency for x-rays reaches a maximum of approximately 10% near photon energies of 1.25 keV. The detection efficiency then falls and levels off at approximately 2 % near photon energies of 0.0125 MeV, remaining essentially constant up to photon energies of 0.125 MeV. Thus, let us assume that the detection efficiencies for 1.274 MeV, 0.511 MeV, and high energy continuum photons are approximately the same and is denoted by $\epsilon$ (the probability is quite low for triplet-state emission of photons less than 50 keV [13]). In addition, because gamma photons possess such high penetration abilities through matter, let us assume that detection efficiencies do not depend on photon angle of incidence (this is a very reasonable assumption since the total thickness of each detector is on the order of 1 mm). The fraction of annihilation events that occur through the singlet channel (emission of two photons) is denoted by $f_1$ and the triplet channel (emission of three photons) by $f_3$. These fractions must satisfy $f_1 + f_3 = 1$. The total count rate on any detector can then be obtained from

$$R_{tot} = \epsilon \frac{\Omega}{4\pi} \left[ 1 + \alpha f_1 + \beta f_3 \right] R_0 \tag{1}$$

where $\Omega$ is the detector solid angle, $R_0$ is the source activity, $\alpha = (2)(.9) = 1.8$, and $\beta = (3)(.9) = 2.7$ (these factors take into account that there are 2 singlet photons and 3 triplet photons that may be emitted in 90% of all decay events). Note the factor in the brackets on the right hand side of Equation 1 has upper and lower limits

$$2.8 \leq \left[ 1 + \alpha f_1 + \beta f_3 \right] \leq 3.7 \ . \tag{2}$$

Thus, total count rates, and subsequent estimates of absolute detector efficiencies, will not be severely dependent on choices of $f_1$ and $f_3$.

The annihilation/annihilation coincidence rate $R_{AA}$ can be written as the start-detector singlet-annihilation detection rate multiplied by the probability of the correlated annihilation photon being detected by the stop detector

$$R_{AA} = \left[ \epsilon_{start} \frac{\Omega_{start}}{4\pi} \alpha f_1 R_0 \right] \epsilon_{stop} \frac{\Omega_{stop}}{\Omega_{start}},$$

$$R_{AA} = \epsilon_{start} \epsilon_{stop} \frac{\Omega_{stop}}{4\pi} \alpha f_1 R_0 \ . \tag{3}$$

In Eq. 3 above, the "start" detector always denotes the detector closest to the source. (The ratio $\Omega_{stop}/\Omega_{start}$ must be less than or equal to one).

The total decay/annihilation rate, for both singlet and triplet states summed together, is given by

$$R_{DA} = 2 \left[ \epsilon_{start} \frac{\Omega_{start}}{4\pi} R_0 \right] \epsilon_{stop} \frac{\Omega_{stop}}{4\pi} (\alpha f_1 + \beta f_3) \tag{4}$$

(the factor of two takes into account the alternative possibility of an annihilation photon striking the "start" detector with the decay photon striking the "stop" detector).

The exact solid angles of each detector are given by

$$\Omega_{start} = 2\pi\left[1 - \frac{s}{\sqrt{s^2 + R^2}}\right],$$

$$\Omega_{stop} = 2\pi\left[1 - \frac{[L-s]}{\sqrt{[L-s]^2 + R^2}}\right], \quad (5)$$

where $s$ is the distance between the source and start detector, $L$ is the distance between both detectors, and $R$ is the active-area radius of each detector.

As a specific example, let us assume that the detector efficiencies are $\epsilon_{start} = \epsilon_{stop} = 0.005$. In addition, let us assume that $f_1 = f_3 = 1/2$. In the current experimental setup, the distance between the detectors is $L = 10$ cm and the active-area radius is $R = 1.25$ cm. The $^{22}$Na source used in this work was initially calibrated at 8.38 $\mu$Ci. Thus, let $R_0 = 310{,}000$ s$^{-1}$. A plot of expected annihilation/annihilation and decay/annihilation coincidence rates as a function of source position "$s$" are given in Figure 3. As one can see, the total coincidence rate is *dominated* by annihilation/annihilation events (from the singlet annihilation channel) when the source is located near the midpoint between the detectors. The maximum expected coincidence rate is $\approx 0.10\,\text{s}^{-1}$, or about six per minute. (Experimental background coincident rates, obtained with the source removed and a detection window of $80\,ns$, are zero).

## IV.  EXPERIMENTAL RESULTS

### A.  Preliminary Results

Estimates of overall experimental timing resolution include electronic, detector transit-time spread, and uncertainties in source/annihilation location. As previously stated in Section **II** , the "*maximum possible*" annihilation-location, or source location, error is 47 ps. Uncertainty due to electronic jitter for the combination of both amplifiers was measured by replacing both detectors with an electronic pulser. The measurements yielded 38 ps for each amplifier circuit. Uncertainties due to electronic walk are 45 ps for each detector circuit (these values are those quoted by the manufacturer). Estimates of detector transit-time spread (TTS) can be obtained by interpolating between the values reported by Kume *et al* [12]. As mentioned earlier, the resultant TTS measurements obtained by Kume *et al.* were 42 ps and 28 ps for MCP's with channel diameters of 12 $\mu$m and 6 $\mu$m, respectively.

Interpolation between these values results in a TTS of 37 ps for a 10 $\mu$m pore diameter detector consisting of one channel plate. Since the detectors utilized in this work are chevron-stacks, an overall estimate of TTS for each detector is thus $\approx$ 52 ps (adding errors in quadrature). A summary of all expected experimental uncertainties is given below:

$$\begin{aligned}
electronic\ jitter \quad & \Delta t_j \approx 38\,ps \quad each \\
electronic\ walk \quad & \Delta t_w \approx 45\,ps \quad each \\
transit-time\ spread \quad & \Delta t_{tts} \approx 52\,ps \quad each \\
source\ location \quad & \Delta t_s \approx 47\,ps\ .
\end{aligned} \qquad (6)$$

The total expected experimental uncertainty is then given by (adding errors in quadrature)

$$\Delta t = \sqrt{2(\Delta t_j)^2 + 2(\Delta t_w)^2 + 2(\Delta t_{tts})^2 + (\Delta t_s)^2}$$

$$\Delta t \approx 120\,ps\ . \qquad (7)$$

Obviously, the above experimental estimate depends on accurate knowledge of electronic walk and detector transit time uncertainties. However, as will be described in Section **V**, these uncertainties can be essentially eliminated through an extrapolation technique.

A standard timing spectrum is illustrated in Figure 4. In this case the source was situated at the midpoint between the detectors. The experimental data was fitted to a Lorentzian curve with each data point weighted by the factor $\sqrt{\sqrt{(x_n - 1)^2}}$ where $x_n$ denotes the number of counts in the $n$ channel. (It is important to note that reduced chi-squares, obtained from fitting data to Gaussian distributions, *far exceed* reduced chi-squares utilizing Lorentzian fits).

Single detector count rates for the measured spectrum, illustrated in Figure 4, were 99 s$^{-1}$ for the start detector and 51 s$^{-1}$ for the stop detector. The total measured coincidence rate was 0.060 s$^{-1}$. Using the above single-detector count rate values in Equations 1 & 5 along with the elapsed time-corrected source activity $R_0 = 274,064\,s^{-1}$ and $f_1 = 0.25$ ($f_3 = 1 - f_1$), results in detector efficiencies of $\epsilon_{start} = 0.0070$ and $\epsilon_{stop} = 0.0036$. Equations 3 & 4 then yield a total expected coincidence rate of 0.054 s$^{-1}$ which is within 10% of the measured rate.

However, one may argue that the above analysis ignores the fact that within metallic solids, such as $^{22}Na$ and the aluminum plates sandwiching the source, positrons may annihilate with electrons "without" forming meta-stable positronium, thereby increasing the number of two-photon emission events. In regard to this argument, the formulation given in Eqs. 1 through 5, can easily be recast such that the partial fraction, "$n$", of annihilation events occurring through direct and "pick-off" annihilation is included. In this case, the fraction of events occurring through the two-photon emission channel increases $f_1 > 0.25$. (The cross section for 3-photon emission within the direct annihilation channel is 1/370 that of 2-photon emission). A direct comparison with the *experimentally determined* coincidence rate yields $n = 0.041$, resulting in $f_1 = 0.28$ and $f_3 = 0.72$. Thus, it appears that a large majority of annihilation events are occurring, not within the sodium source or aluminum plates, but within the encapsulating 5 mm thick plastic. (Hundreds of "bubbles", acting as voids, can be seen by the naked eye within the material surrounding the source). However, as previously stated, coincidence detection is *dominated* by two-photon emission events since the detectors are situated 180° from each other.

The absolute timing position of the peak was also checked utilizing an external pulser. The pulser output was attached directly to each anode plate. The pulser test indicated that the annihilation peak was within 700 ps of where it was expected. (It is important to note that the temporal pulse width of signals from MCP detectors are $\approx$ 1 ns).

As a further test to confirm that the measured timing peak was indeed actual detection of annihilation events, the radioactive source was moved and spectra were retaken to look for corresponding movements in the timing position of the peak. Let us assume that an annihilation event occurs at $t = 0$. The detection of annihilation photons at the start and stop inputs of the PTA then occur at

$$t_{start} = \frac{s}{c} + \tau_{start}$$

$$t_{stop} = \frac{[L - s]}{c} + \tau_{stop} + \tau_{delay} \tag{8}$$

where $\tau_{delay}$ is the delay setting on the DB 463 delay box (see Fig. 1), $c$ is the speed of light, and $\tau_{start}$ and $\tau_{stop}$ denote all other existing known or unknown electronic delays within the start and stop circuits. The centroid of the timing peak $T_c$ obtained by the PTA, is then

given by

$$T_c = t_{stop} - t_{start} = \frac{[L - 2s]}{c} + [\tau_{stop} - \tau_{start}] + \tau_{delay} \,. \tag{9}$$

A change in source position $\Delta s$ will then result in the following change in centroid position

$$\Delta T_c = \frac{-2\,\Delta s}{c} \,, \tag{10}$$

which is **independent of any electronic settings or delays**.

Figure 5 illustrates the results obtained by moving the source from an initial position of $s = 2.6$ cm to $s = 6.3$ cm. The expected timing shift is therefore $\Delta T_c = -247$ ps. The measured time shift, obtained from the actual centroid positions, is $-286$ ps. Note that the discrepancy between these values (39 ps) may be accounted for by the inherent asymmetries in the spectra.

Further experimental measurements were performed to try and find possible causes of the observed asymmetry. The spectral asymmetry was found to have a strong dependence on the source position. When the source was situated close to the start detector ($s < L/2$) the asymmetry favored later arrival times. When the source was positioned close to the stop detector ($s > L/2$) the asymmetry favored earlier arrival times. When the source was situated at midpoint ($s = L/2$) spectra exhibited a more symmetric shape.

At first glance, one may attribute the observed asymmetry to decay/annihilation coincidences. As previously stated, in 90% of all decay events a prompt 1.274 MeV decay photon is followed by the emission of a positron. Due to positron thermalization times, subsequent annihilation photons are not always emitted simultaneously with the decay photon, but at a later time. The late arrivals of annihilation photons would thus generate an asymmetry in the timing spectrum favoring later arrival times (assuming the decay photon initiates the start signal). However, it is also possible to obtain a start signal from an annihilation photon, with the stop signal initiated by a decay photon. In this case, the stop signal would arrive *before* the start signal, leading to an asymmetry favoring earlier times. (One must remember that the stop signal is always delayed more than 50 ns due to electronic dead time). Since the probability is the same for these coincidence events, it necessarily follows that timing spectra should always be "symmetric", regardless of source position.

Additional tests involving increasing the thickness of aluminum surrounding the $^{22}$Na source also failed to make any changes in spectral asymmetry. (This ruled out the possibility of positrons escaping from the source).

We then focussed our attention on possible electronic causes. Electronic tests and measurements, including adjustments of walk settings, impedance matching, etc., failed to eliminate the observed asymmetries. However, examinations of pulse height distributions, while not yielding any explanation of the asymmetries, did enable us to dramatically improve the FWHM timing resolution.

### B. Electronic Tagging of Over-Range Pulses

Microchannel plate detectors are typically utilized for photons and/or charged particles with energies from the ultraviolet range to a few keV. For the sake of simplicity, let us define "mid-energy" electrons as electrons possessing 1 keV of energy which have the highest MCP detection efficiency of approximately 50 %. Figure 6 gives a qualitative comparison of the pulse height distribution of mid-energy electrons with that of high energy gamma photons. (Note: at present, our laboratory is not fully equipped with state-of-the-art electronic equipment such as a digital storage oscilloscope, pulse height analyzer, etc., needed to obtain *absolute* measurements of pulse height distributions. However, the gamma photon pulse height distribution illustrated in Figure 6 is based on relative measurements obtained utilizing antiquated $\mu$sec based electronics. In addition, measured count rates are observed to vary exponentially with threshold settings, which conclusively proves that the pulse height distributions are exponential in character. The pulse height distribution for mid-energy electrons is obtained from experimental data supplied directly by the detector manufacturer, Burle Industries).

The 9327 amp/disc is designed to accept pulses with amplitudes up to 30 mV. Over-ranged pulses, with amplitudes > 30 mV, will drive the 9327 amplifier into saturation. Obviously, the constant-fraction (zero-crossing) circuitry cannot reliably trigger timing pulses for over-ranged signals [14]. For mid-energy electrons, over-ranged pulses have a negligible effect since the fraction of over-ranged detection events is only a few percent.

In contrast, it was discovered that the fraction of over-ranged events for gamma photons were as high as 30 to 40 % which has a substantial effect on timing resolution. Unfortunately, due to the exponential character of the gamma-photon pulse-height distribution (Fig. 6), reducing the detector or amplifier gain only decreases overall count rates and *does not significantly reduce the fraction* of over-ranged pulses. A better method is to utilize as much detector or amplifier gain as possible, while eliminating the over-ranging signals by electronically "tagging" all pulses exceeding 30 mV.

Figure 7 illustrates the method utilized for electronic tagging of over-range pulses. The 9327 amp/disc provides an "amplifier output" for direct monitoring of the input signal. The amp out signals are passed into separate constant fraction discriminators (584 and 473 CF discriminators, which happened to be available) operating in the "leading edge" trigger mode. Thresholds on the 584 and 473 CF Disc's are set to trigger exactly when the input pulses entering the 9327's exceed 30 mV. The 584 and 473 CF Disc's provide positive NIM logic output pulses (which are TTL logic compatible) for electronic tagging.

The 9308 PTA is operated in the "list" mode. In this mode, the PTA generates a sequential "list" file with each line containing a start/stop time interval along with a tag number for each event. The PTA provides four binary tag inputs (TTL logic) to generate a corresponding tag number, ranging from zero to 15, for event identification. (Tag inputs must arrive within 30 ns after each stop event). In addition, since the PTA does not have to generate a real-time histogram, it frees up more internal memory, which increases the instrumental timing resolution to 0.305 ps. Thus, utilizing the PTA in the list mode along with electronic tagging of events, each start/stop measurement obtained includes a tag number identifying whether that particular event had any start and/or stop over-ranging pulses.

Figure 8 illustrates typical results obtained utilizing electronic tagging of over-ranged pulses. As Figure 8 indicates, when the stop signal over-ranges (with start signal *not* over-ranging) the stop signal tends to arrive at a later time. Start signals that over-range also tend to arrive late. This can be seen in the start-overange spectrum which shows the peak arriving *earlier* in time. As Figure 8 clearly illustrates, the combined effect of start/stop over-ranging is an undesirable broadening of the overall FWHM of the centroid in the timing spectrum.

Unfortunately, over-range tagging does not eliminate the previously observed asymmetries. Figure 9 compares non-over-ranged timing spectra when the source is moved from $s = 6.4$ cm to $s = 4.9$ cm. The expected time shift of the centroid (Equation 8) is $+100$ ps. The measured shift is $+118$ ps. The percent discrepancy between measured and expected shift values (18 ps/118 ps) basically remains unchanged from previous measurements.

In order to examine effects on the overall timing resolution, the $^{22}$Na source was repositioned back to the midpoint location ($s = L/2$) and a timing spectrum was taken. Figure 10 illustrates the results obtained with over-ranged pulses included. Figure 11 illustrates the same spectrum as in Figure 10, but without over-ranged pulses. As one can see, the FWHM was substantially reduced. We would like to point out that the FWHM reported in Fig. 11 is an overestimate. A better reduced chi-square can be obtained by fitting the data ranging only from 103.5 ns to 104.5 ns, and the results are illustrated in Fig. 12. As Figure 12 indicates, a timing resolution of 166 ps FWHM is obtained with weighted reduced chi-square of 1.005.

Thus, the method of electronic over-range tagging has proven to be successful. In comparison with the FWHM illustrated in Fig. 4, this method has improved the overall timing resolution by $\approx$ 155 ps (adding all errors in quadrature). However, the experimental time resolution **can still be further improved** by electronic selection of detector-transit or "electron-avalanche propagation" time.

### C. Electronic Selection of Electron-Avalanche Propagation Time

As described in the previous section, the exponential character of the gamma-photon pulse-height distribution (PHD) (Fig. 6) necessitated the use of electronic tagging to eliminate over-ranged pulses. However, in regard to these distributions, a fundamental question arises. Why are the pulse height distributions for mid-energy electrons and gamma photons so different?

The differences in the pulse height distributions can be explained, in a qualitative fashion, by examining penetration depths as illustrated in Figure 13. Assuming that medium energy electrons are incident normal to the front surface of the MCP, the maximum depth at which the electrons can penetrate inside the micro-channel $d_{max}$ is given by

$$d_{max} = d_c \cot \theta \, , \tag{11}$$

where $d_c$ is the channel (pore) diameter and $\theta$ is the detector bias angle. As an example, the MCP's used in this work have a 10 $\mu$m pore diameter and 8$^o$ bias angle. The maximum depth of penetration is therefore 71 $\mu$m, which is approximately 18% of the thickness of a single channel plate. Gamma photons, on the other hand, have no maximum penetration depth.

The amplitude of the anode pulse, generated by either a single electron or photon, depends upon the number of electrons produced in the avalanche. A photon or electron striking near the face of the MCP (surface farthest from the anode) will produce the largest electron avalanche and subsequently the largest pulse on the anode. A photon striking near the base of the MCP (surface closest to the anode) will produce the smallest electron avalanche and subsequently the smallest pulse on the anode. Thus, in a qualitative sense, the position where the particle strikes the inner channel wall and the avalanche starts, measured relative to the anode, is proportional to the pulse amplitude.

Therefore, the pulse height distributions illustrated in Fig. 6 can also be viewed as a plot of the number of pulses generated versus avalanche creation position, by replacing the pulse amplitude with the distance that the particle strikes the detector, measured relative to the anode plate. Since very few electrons are able to penetrate further than $d_{max}$ in the detector, a sharp gaussian peak occurs in the electron PHD. Gamma photons, on the other hand, are highly penetrating. The number of gamma photons that pass freely through matter is expressed by a decreasing exponential function. Thus, it is not surprising that the gamma-photon PHD also exhibits a corresponding exponential type of behavior.

There also exists, again in a qualitative sense, a direct relation between pulse height and electron-avalanche propagation time (detector transit time). Since a large pulse corresponds to an avalanche being created far from the anode, the propagation time of this avalanche will be longer than that of a smaller pulse, since the smaller pulse corresponds to the avalanche being created closer to the anode. Thus, the pulse height distributions illustrated in Fig. 6 can also be viewed as a plot of the number of pulses versus avalanche propagation time (replace pulse amplitude with propagation time).

This qualitative model, however, may also be described in a fully quantitative fashion through derivation of the expected pulse height distribution for incident gamma photons.

### D. Derivation of pulse height distribution

The total width of the chevron stack is denoted by $L$. Let us assume that the electron avalanche is first generated at position $x$ measured from the front face of the MCP (face farthest from the anode). The probability of the avalanche starting between $x$ and $x + dx$ is $P(x)dx$. The probability of generating a pulse amplitude between $V$ and $V + dV$ is given by $G(V)dV$. The amplitude $V$ of the generated pulse depends upon $x$ and is denoted by $V(x)$. Utilizing the fundamental theorem of probability, the pulse height distribution $G(V)$ can then be obtained from

$$G(V) = P(x)\frac{dx}{dV} . \qquad (12)$$

The pulse amplitude, as a function of avalanche start position $V(x)$, can be obtained in the following manner. Let $n_e$ denote the average number of electrons ejected per collision per electron, and $s_o$ denote the average electron propagation distance (along $x$) between collisions. The pulse amplitude is then given by

$$V(x) = \frac{e}{c} n_e^{(L-x)/s_o} , \qquad (13)$$

where $e$ is the charge on the electron and $c$ is the detector/anode capacitance ($c = \epsilon_o A/d \approx 5\,pF$). (Although Eq. 13 was initially obtained through a geometrical series type argument, a full derivation based on physical principles was first reported by Adams and Manley in 1966 [15]). As an example, utilizing the values $s_o = L/44$ and $n_e = 1.4$ results in a maximum pulse amplitude (for $x = 0$) of $V_{max} = 86\,mV$. This corresponds to a maximum of 2.7 x $10^6$ electrons generated in the avalanche.

Equation 13 can now be solved for $x$

$$x(V) = L - \frac{s_o}{\ln(n_e)} \ln(\frac{cV}{e}) . \qquad (14)$$

The derivative of $x(V)$ is then

$$\frac{dx}{dV} = -\frac{s_o}{\ln(n_e)} \frac{1}{V} . \qquad (15)$$

Since the detector width $L$ has such a miniscule effect on stopping the incident gamma photons, the probability function $P(x)$ is *essentially constant* over the width of the detector ($P(x) \approx A_o$). In addition, a positive change in $x$ results in a negative change in $V$ ($dx = -dV$). Thus, the pulse height distribution is given by

$$G(V) = -A_o \frac{dx}{dV} = \frac{A_o s_o}{\ln(n_e)} \frac{1}{V} . \qquad (16)$$

As stated earlier, due to the lack of proper state-of-the-art electronic equipment, we could not obtain absolute measurements of pulse height distributions. However, we could do the next best thing. We measured the total count rate on the detector versus amplifier threshold setting $V_{th}$. This measured count rate $C(V_{th})$ can be directly obtained from Equation 16 (within an arbitrary constant $B_o$)

$$C(V_{th}) = B_o \int_{V_{th}}^{V_{max}} G(V) dV \tag{17}$$

$$C(V_{th}) = \frac{B_o A_o s_o}{\ln(n_e)} [\ln(V_{max}) - \ln(V_{th})] \tag{18}$$

$$C(V_{th}) = C_o [D_o - \ln(V_{th})] \tag{19}$$

(note: $V_{max} = (e/c) n_e^{L/s_o}$). A comparison of Equation 19 with the experimental data is presented in Figure 14 for $C_o = 656$ and $D_o = 6.87$. As can be seen in Figure 14, Equation 19 is in *excellent* agreement with the data.

The agreement of Equation 19 with the experimental data implies that our choice of $V(x)$ is correct. It then follows that a finite amplitude selection window $\Delta V$, centered at $V$, necessarily restricts avalanche start positions within a range $\Delta x$ given by

$$|\Delta x| = \frac{s_o}{\ln(n_e)} \ln(\frac{V + \Delta V}{V}) . \tag{20}$$

In the limit as $\Delta V \to 0$, $\Delta x \to 0$. Since the detector transit time $t_{trans}$ is directly related to $x$, it then follows that $\Delta t_{tts}$ tends to zero (or an extremely small value) as $\Delta V \to 0$. (In addition, the number of electrons collected in the avalanche approaches a fixed value).

Since the avalanche propagation time is proportional to pulse amplitude, it becomes apparent that by decreasing the amplitude "selection window", set by the threshold and over-range settings (Fig. 6), one can electronically "select" a particular range in electron avalanche propagation times. For the case of incident electrons, this could be centered around the peak in the Gaussian shaped PHD. In contrast, for the case of gamma photons, it is better to keep the threshold as low as possible while decreasing the over-range value in order to preserve count rates.

### E. Experimental Test of Electronic Selection Method

To test this model, the over-range threshold settings on the 584 and 473 CF Disc's were lowered and a timing spectrum was accumulated with the source positioned at midpoint ($s = L/2$). In this case, however, not only were over-ranged pulses "tagged" but all other pulses outside the avalanche-propagation time window were also tagged. The 584/473 over-range thresholds were set such that the count rate of tagged pulses on both start and stop signals comprised approximately 75 % of respective total count rates.

Figure 15 illustrates the final PTA spectrum obtained. As one can see, an experimental FWHM of **120 ps** was easily achieved. It is important to note that this experimental time resolution is *less* than the lifetime of positronium atoms in vacuum.[16, 17]

However, due to the weighting factor $\sqrt{\sqrt{(x_n - 1)^2}}$ utilized in the fitting routine, this timing resolution is somewhat underestimated. Thus, an additional Lorentzian fit was made to the data presented in Figure 15, without using any weighting factors, and resulted in a FWHM = 155 ps. Additional experimental runs were also taken, after further reducing the over-range thresholds, and one such run is illustrated in Figure 16. In this case, the 584/473 over-range thresholds were set such that the count rate of tagged pulses on both start and stop signals comprised approximately 80 % of respective total count rates. The FWHM of 138 ps reported in Fig. 16 was obtained without any weighting factors.

Thus, the method of "electronic selection" of transit or avalanche propagation times has proven to be successful. Before reducing the over-range thresholds, the FWHM from Fig. 12 was 166 ps, utilizing a weighted fit. In order to compare with the results in Fig. 16, a "non-weighted" fit of the data in Fig. 12 results in a FWHM of 191 ps. Thus, the transit time selection method actually reduced the FWHM from 191 ps to 138 ps.

## V. FINAL RESULTS

An experimental estimate of the *minimum quantum time uncertainty*, between detection of the two annihilation photons, can be obtained through extrapolation of the known variance of the FWHM with decreasing over-range thresholds.

Experimentally determined FWHM's obtained from further additional measurements, without using weighting factors, are plotted versus the percentile of non-tagged count rates in Fig. 17. The total expected measured FWHM is given by:

$$\Delta t = \sqrt{(\Delta t_{QM})^2 + 2(\Delta t_w)^2 + 2(\Delta t_{tts})^2 + 2(\Delta t_j)^2 + (\Delta t_s)^2}. \qquad (21)$$

As one reduces non-tagged count rates, by lowering the upper thresholds, **both** amplitude-dependent electronic walk and transit-time spread are also reduced. Thus, the combined uncertainty in electronic walk and transit-time ($\sqrt{2(\Delta t_w)^2 + 2(\Delta t_{tts})^2}$) is nominally proportional to the percentile of non-tagged count rates. Since $\Delta t_j$ and $\Delta t_s$ have been experimentally determined (see Eq. 6), the following equation can then be fitted to the experimental data in Fig. 17

$$\Delta t = \sqrt{(a)^2 + (b\,x)^2 + 2(\Delta t_j)^2 + (\Delta t_s)^2}, \qquad (22)$$

where the fitting parameters are $a$ and $b$, and $x$ denotes the total percentile "non-tagged" count rate

$$x = \sqrt{(R_{start})^2 + (R_{stop})^2}, \qquad (23)$$

where $R_{start}$ and $R_{stop}$ are the individual percentile non-tagged count rates for the start and stop channels, respectively (which are obtained directly from the PTA list data). The final quantum uncertainty in time is given by the parameter $a$. Fitting Eq. 22 to the data in Fig. 17, results in $\Delta t_{QM} = 123 \pm 22\,ps$. The agreement between this value and the lifetime of singlet positronium atoms in vacuum is intriguing.

## VI. CONCLUSIONS

The concept and subsequent experimental verification of the proportionality between pulse amplitude and detector transit time has considerably improved the overall timing resolution for detection of high energy gamma photons utilizing microchannel plate detectors. The FWHM was initially reduced by $\approx$ 155 ps through utilization of electronic tagging of over-ranged pulses. A final reduction of $\approx$ 132 ps was achieved through both the reduction in electronic walk and reduction in electron-avalanche propagation time spread, resulting in a final FWHM of ≈ 138 ps. In addition, extrapolation of the known variance of the FWHM with decreasing over-range thresholds results in an absolute minimum uncertainty in detection time of $123 \pm 22\, ps$ between the arrival of the two annihilation photons.

Our experimental measurements reported here give strong evidence *against* instantaneous spatial-localization of γ-photons due to measurement-induced non-local quantum wavefunction collapse. Although the results disagree with earlier predictions of Einstein, Podolsky, and Rosen [1], they are in agreement with a recent quantum-mechanical analysis of Irby [9]. The analysis of Irby is essentially the same as that first presented by Einstein, Podolsky, and Rosen. The main difference, however, is that Irby's analysis includes both time dependence and conservation of energy.

The experimental results are also the first that imply momentum is conserved only after the quantum uncertainty in time has elapsed [18].

## VII.    ACKNOWLEDGMENTS

This work was entirely supported by internal funds from the University of South Alabama and the University of South Alabama Research Council (USARC). The author also thanks Justin Sanders for enlightening discussions and loan of the 473 constant fraction discriminator, and Paul Helminger for his insight and help in our final time-resolution data analysis.

[1] A. Einstein, B. Podolsky, and N. Rosen, Phys. Rev. **47**, 777, (1935).

[2] M. H. Rubin, D. N. Klyshko, Y. H. Shih, and A. V. Sergienko, Phys. Rev. A, **50**, 5122, (1994).

[3] J. S. Bell, Physics 1 (Long Island City, N.Y.)1964, reprinted in J. S. Bell, "Speakable and unspeakable in quantum mechanics", (Cambridge University, Cambridge) 1987.

[4] A. Aspect, J. Dalibard, and G. Roger, Phys. Rev. Lett., **49**, 1804 (1982).

[5] J. D. Franson, Phys. Rev. A, **44**, 4552 (1991).

[6] J. D. Franson, Phys. Rev. Lett. **62**, 2205 (1989).

[7] C. K. Hong, Z. Y. Ou, and L. Mandel, Phys. Rev. Lett. **59**, 2044 (1987).

[8] R. A. Bertlmann and A. Zeilinger, Eds. "Quantum Unspeakables, From Bell to Quantum Information", (Springer-Verlag, New York) 2002.

[9] V. D. Irby, Phys. Rev. A **67**, 034102, (2003).

[10] M. Ito, H. Kume, and K. Oba, IEEE Trans. Nucl. Sci. **NS-31**, 408 (1984).

[11] B. K. F. Young, R. E. Stewart, and J. G. Woodworth, Rev. Sci. Instrum. **57**,2729 (1986).

[12] H. Kume, K. Koyama, K. Nakatsugawa, S. Suzuki, and D. Fatlowitz, Applied Optics **27**, 1170 (1988).

[13] A. Ore and J. L. Powell, Phys. Rev. **75**, 1696 (1949).

[14] D. A. Gedcke and W. J. McDonald, Nucl. Instr. and Meth. **58**(2), 253 (1968).

[15] J. Adams and B. W. Manley, IEEE Trans. Nuc. Sci. , NS-13, 88, (1966).

[16] J. J. Sakurai, in *Advanced Quantum Mechanics*, (Addison-Wesley, Reading Mass. 1977), pg. 216.

[17] M. Charlton and J. W. Humberston, in Positron Physics (Cambridge University Press, Cambridge, 2001), p.264.

[18] H. Yukawa, Proc. Phys. -Math. Soc. Japan, **17**, 48 (1935).

**FIGURE 1.** A simplified schematic illustrating the detection geometry and electronic timing circuitry. The $^{22}$Na source is located at position "s" measured relative to the start MCP. The distance between detectors is L=10 cm.

**FIGURE 2.** Electrical schematic illustrating detection and pick-off circuits. The resistor connected in series with the HV supply is chosen such that the potential at the detector base is $\approx$ -200 V for a HV setting of -2000 V. The specific value required for each detector depends on the resistance of that specific detector.

**FIGURE 3.** Expected annihilation/annihilation $R_{AA}$ (long dashed line) and decay/annihilation $R_{DA}$ (short dashed line) coincidence rates are plotted as a function of source position "s". $R_{DA}$ coincidence rate includes annihilation photons from both singlet and triplet states, whereas $R_{AA}$ involves only singlet states. The total coincidence rate $R_{total} = R_{AA} + R_{DA}$ is given by the solid line.

**FIGURE 4.** Experimental PTA spectrum for source situated at midpoint ($s = L/2$). The full-width half-maximum (FWHM) was obtained by fitting a Lorentzian curve to the data (see text). $\chi^2$ denotes the weighted reduced chi-square. Note that the absolute value of the centroid position depends on external delays and is irrelevant to the FWHM.

**FIGURE 5.** Experimental PTA spectra illustrating the dependence of the timing peak on source position. Source, initially located at $s = 2.6$ cm, was moved to $s = 6.3$ cm.

**FIGURE 6.** A qualitative illustration of detector pulse height distributions (PHD). The solid line represents a PHD for mid-energy ($\approx$ 1 keV) electrons. The dashed line represents the PHD for 0.511 MeV gamma photons.

**FIGURE 7.** Electrical schematic illustrating method utilized for electronic "tagging" of over-ranged pulses (see text).

**FIGURE 8.** Experimental results utilizing over-range tagging. The lower most plot illustrates the full spectrum which includes over-ranged pulses.

**FIGURE 9.** Experimental PTA spectra illustrating the dependence of the timing peak on source position. In this case, all over-ranged events have been removed from the spectrum.

**FIGURE 10.**  Experimental PTA spectrum, including all over-ranged events, for source located at the midpoint location $s = L/2$.

**FIGURE 11.**  Experimental PTA spectrum, excluding all over-ranged events (obtained from spectrum illustrated in Fig. 10).

**FIGURE 12.**  Experimental spectrum from Fig. 11 is plotted ranging only from 103.5 ns to 104.5 ns. A Lorentzian function is then fitted to the data.

**FIGURE 13.**  Schematic diagram illustrating penetration depths of mid-energy electrons and gamma photons (see text).

**FIGURE 14.**  Detector count rate is plotted versus threshold setting. Experimental data is represented by solid circles. Solid line is a fit of Eq. 19 to the data.

**FIGURE 15.**  Final high resolution PTA spectrum. The timing spectrum was obtained utilizing the method of electronic selection of electron-avalanche propagation times (see text).

**FIGURE 16.**  Final high resolution PTA spectrum obtained without utilizing weighted fit (see text).

**FIGURE 17.**  Experimental FWHM measurements are plotted versus percentile non-tagged count rate (see text). The solid line is the fit of Eq. 22 to the experimental data. Individual error bars represent 3 standard deviations. The quoted overall error is based on a 95 % confidence interval.

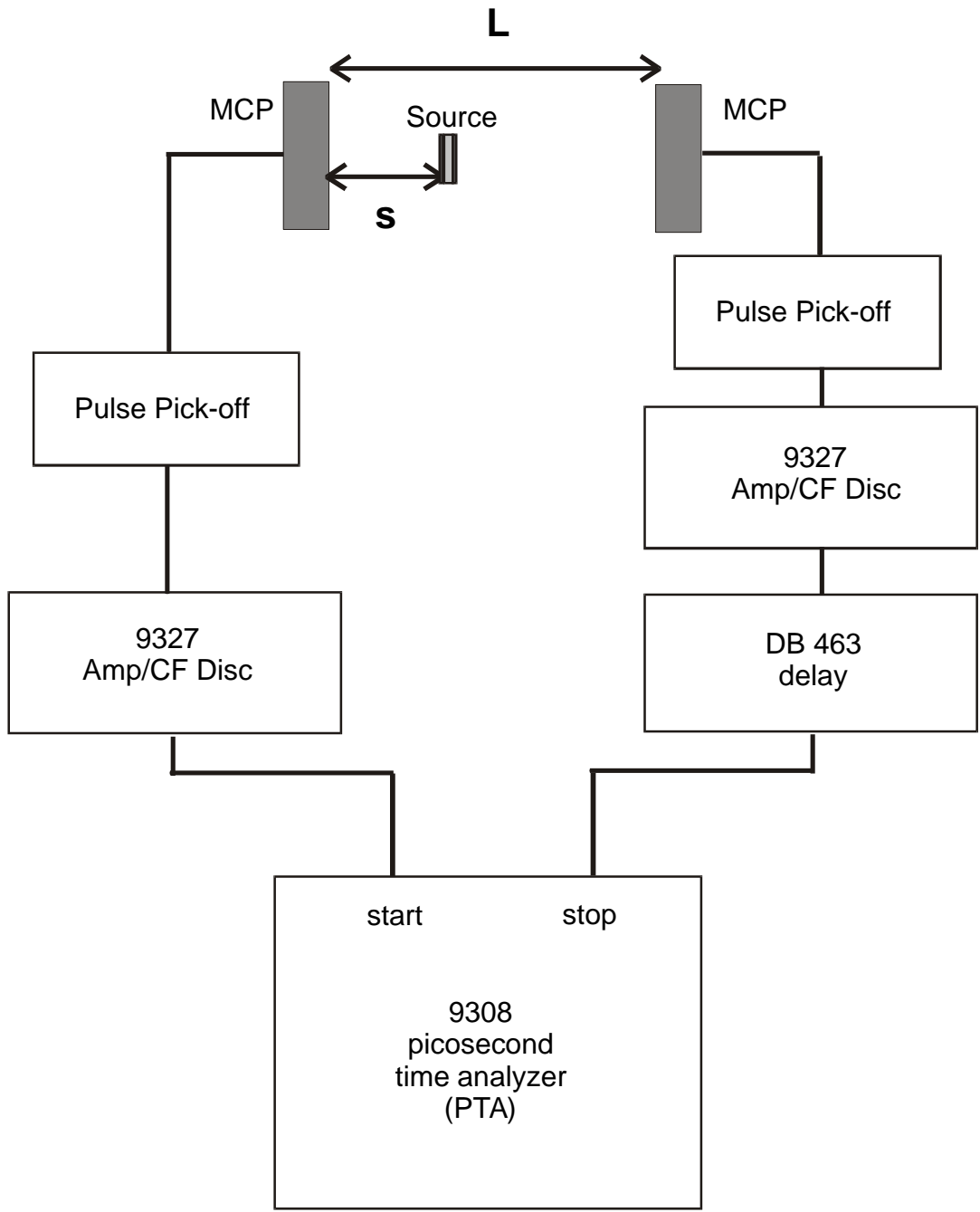

**Figure 1**

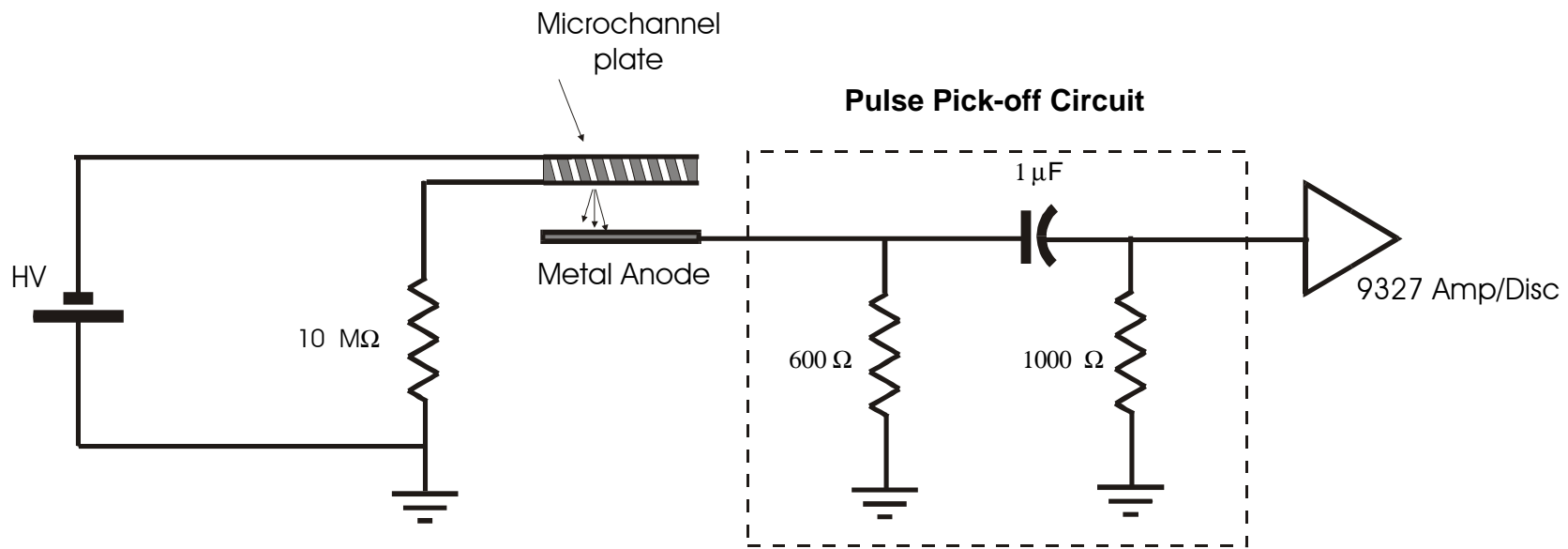

**Figure 2**

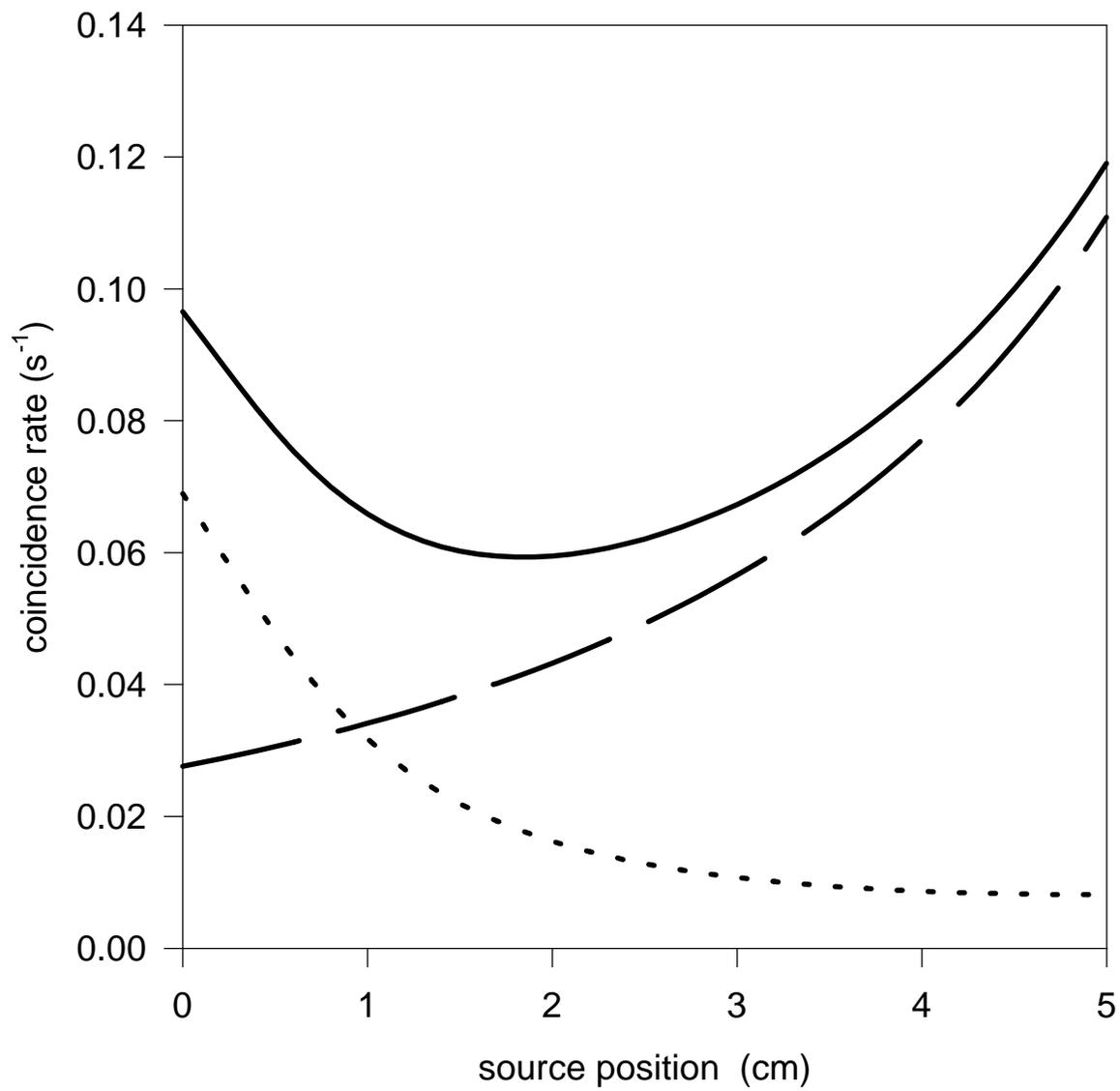

**Figure 3**

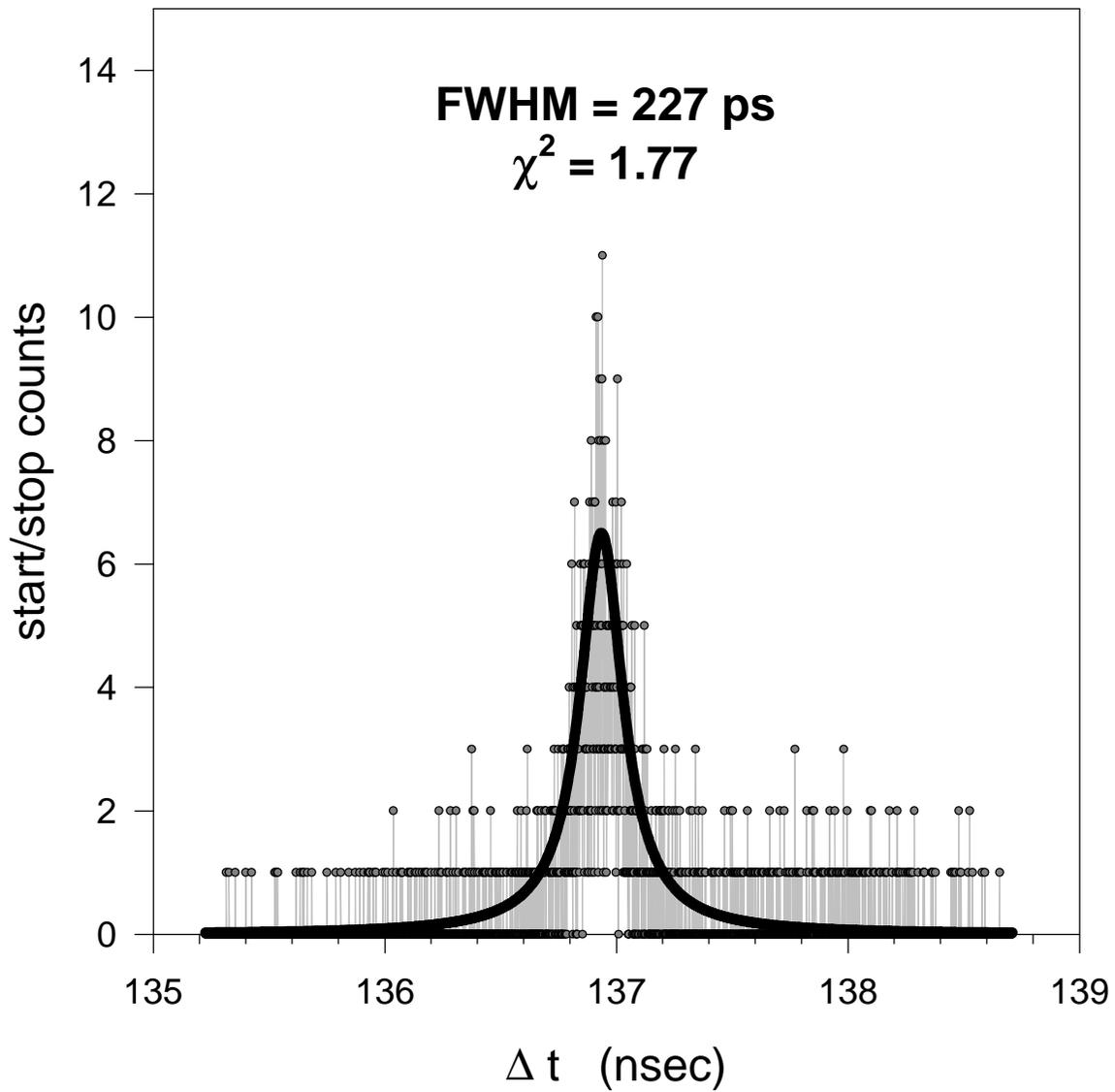

**Figure 4**

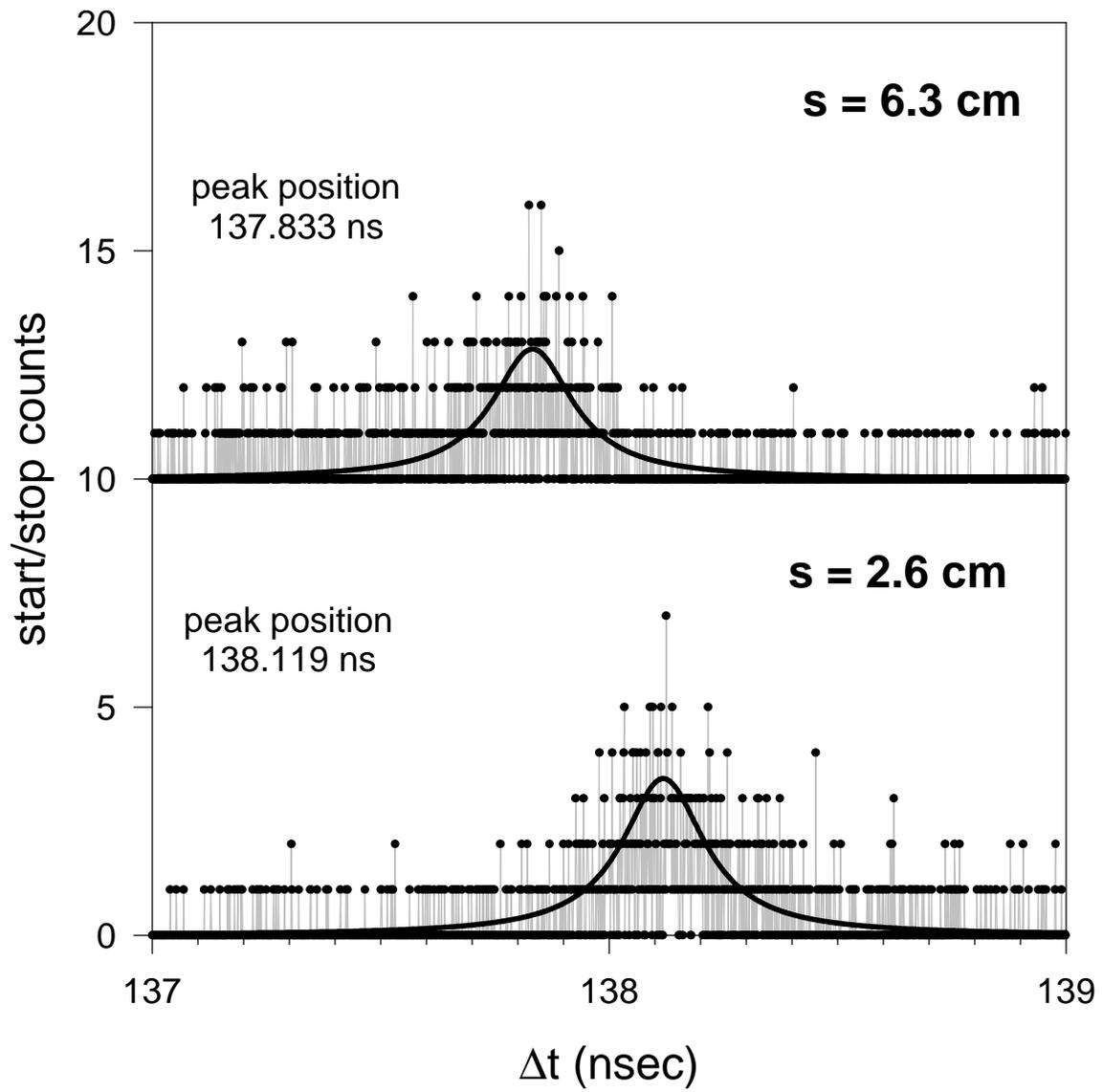

**Figure 5**

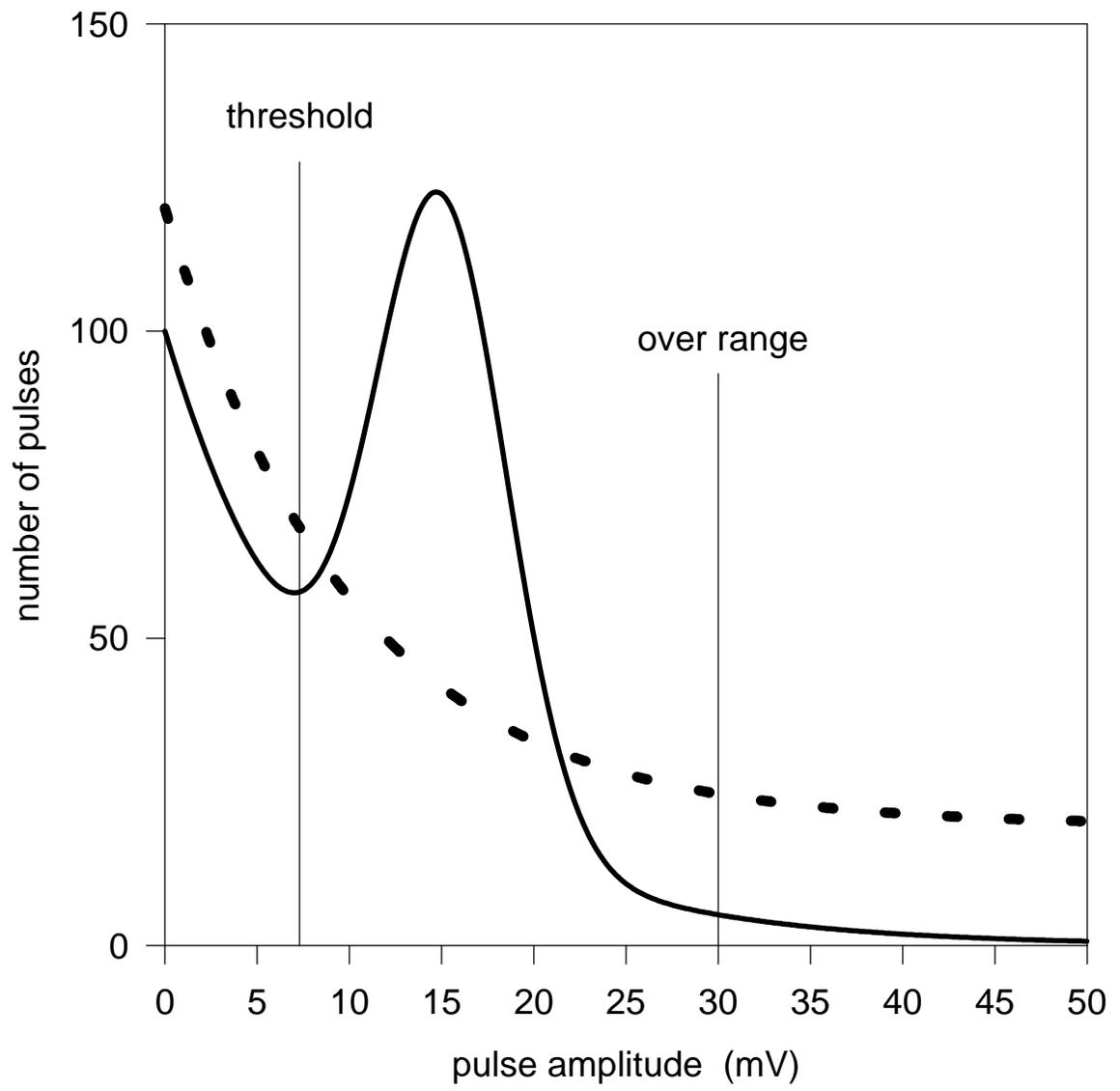

**Figure 6**

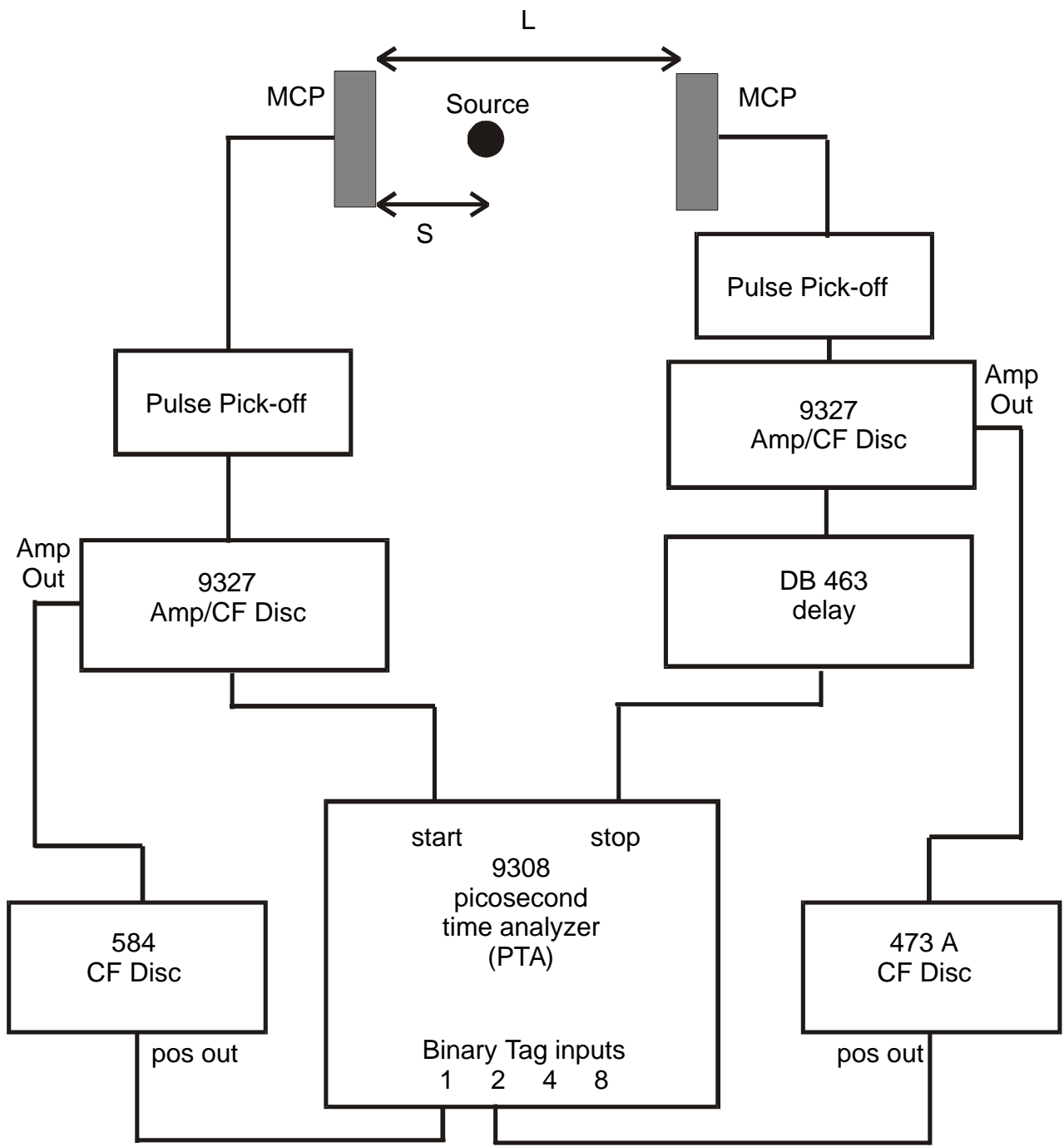

Figure 7

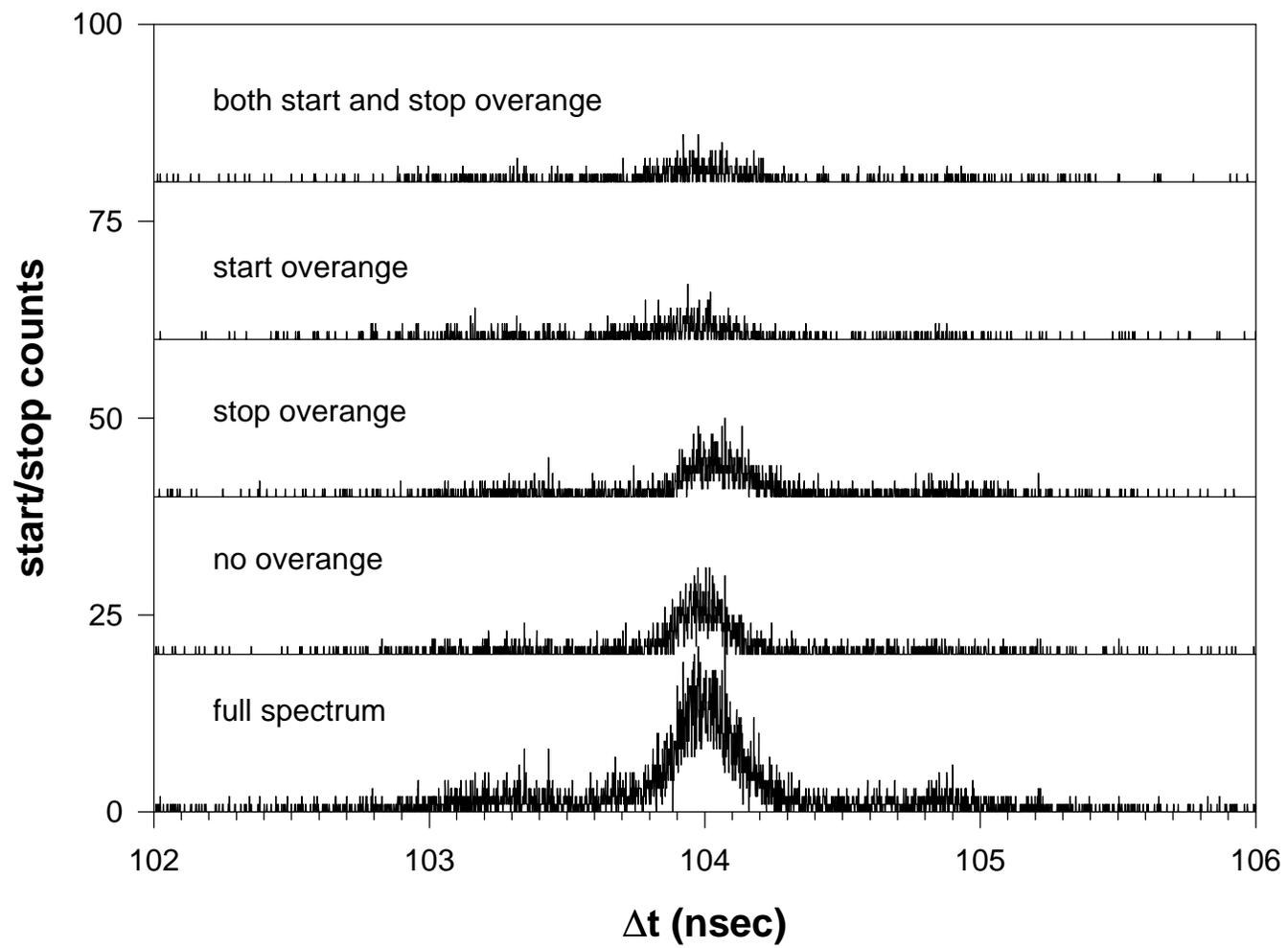

**Figure 8**

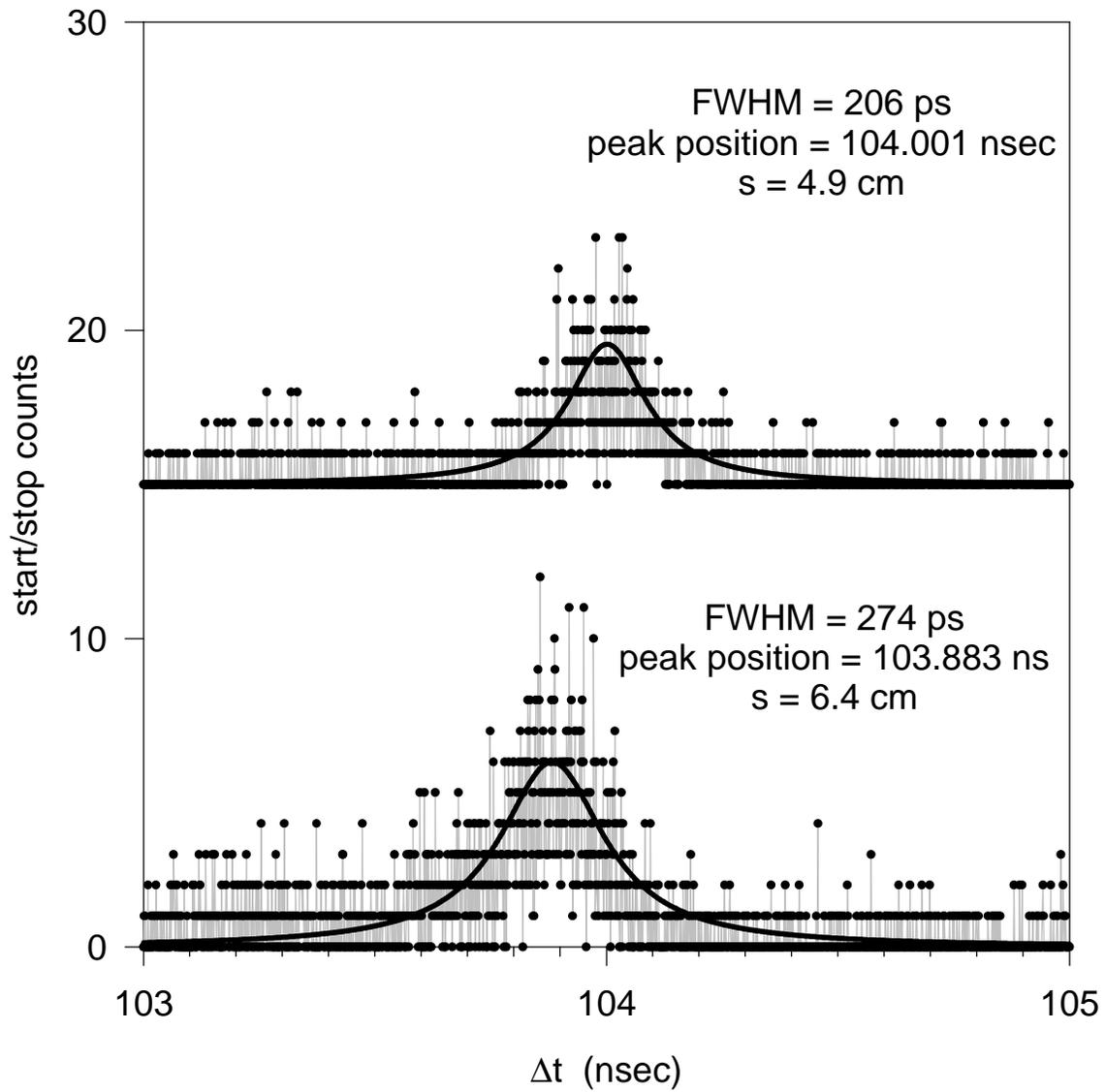

**Figure 9**

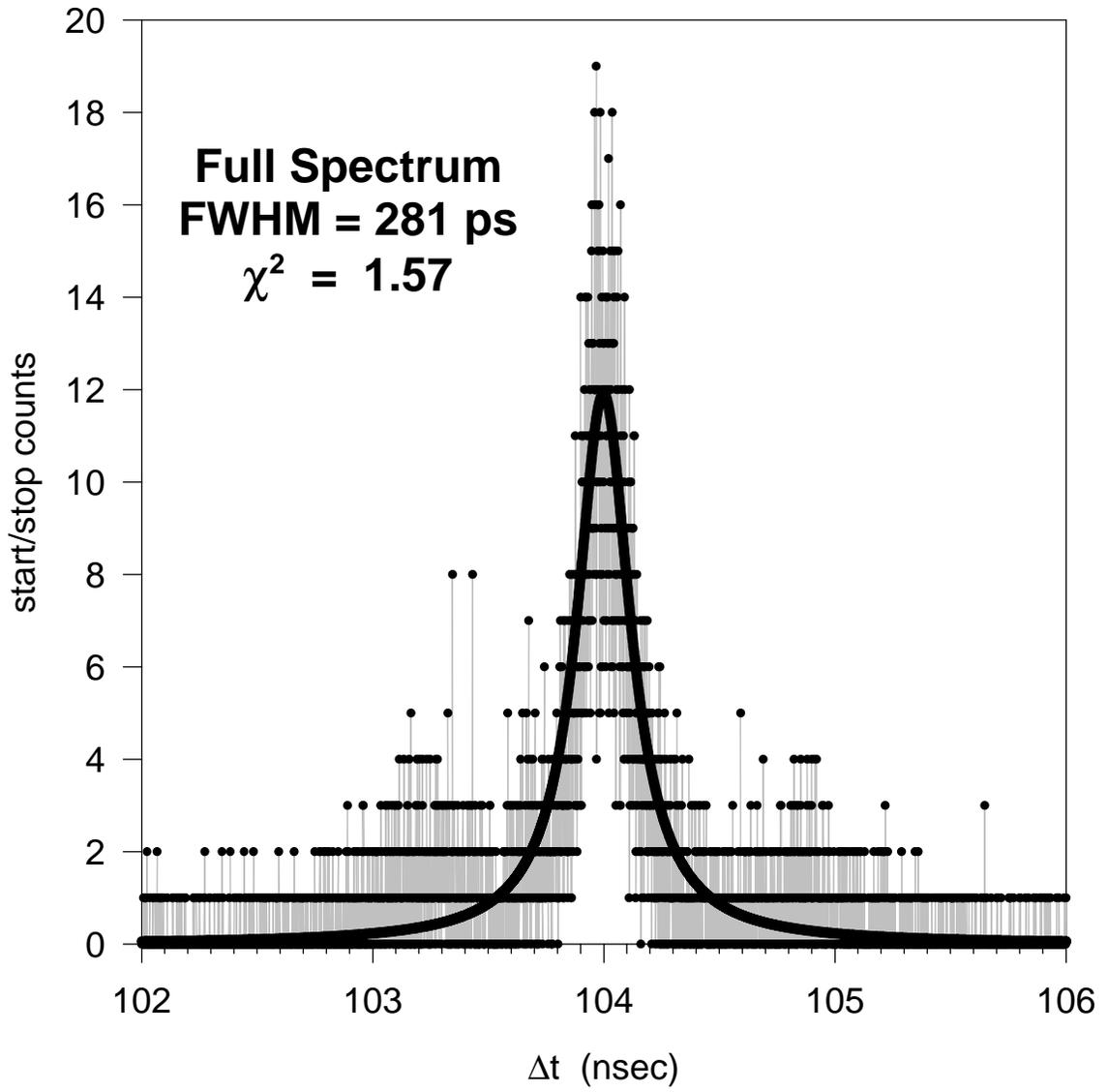

**Figure 10**

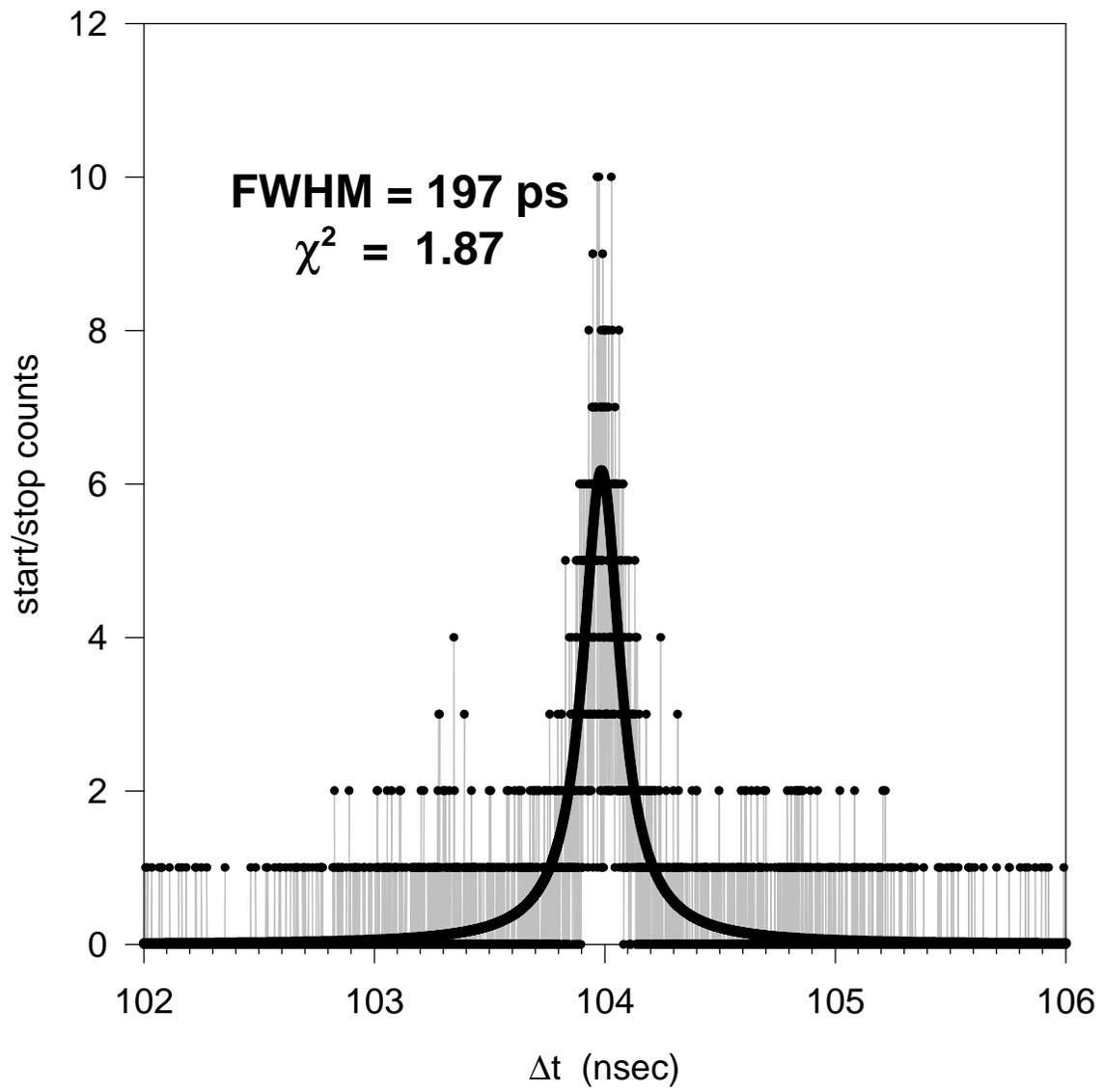

**Figure 11**

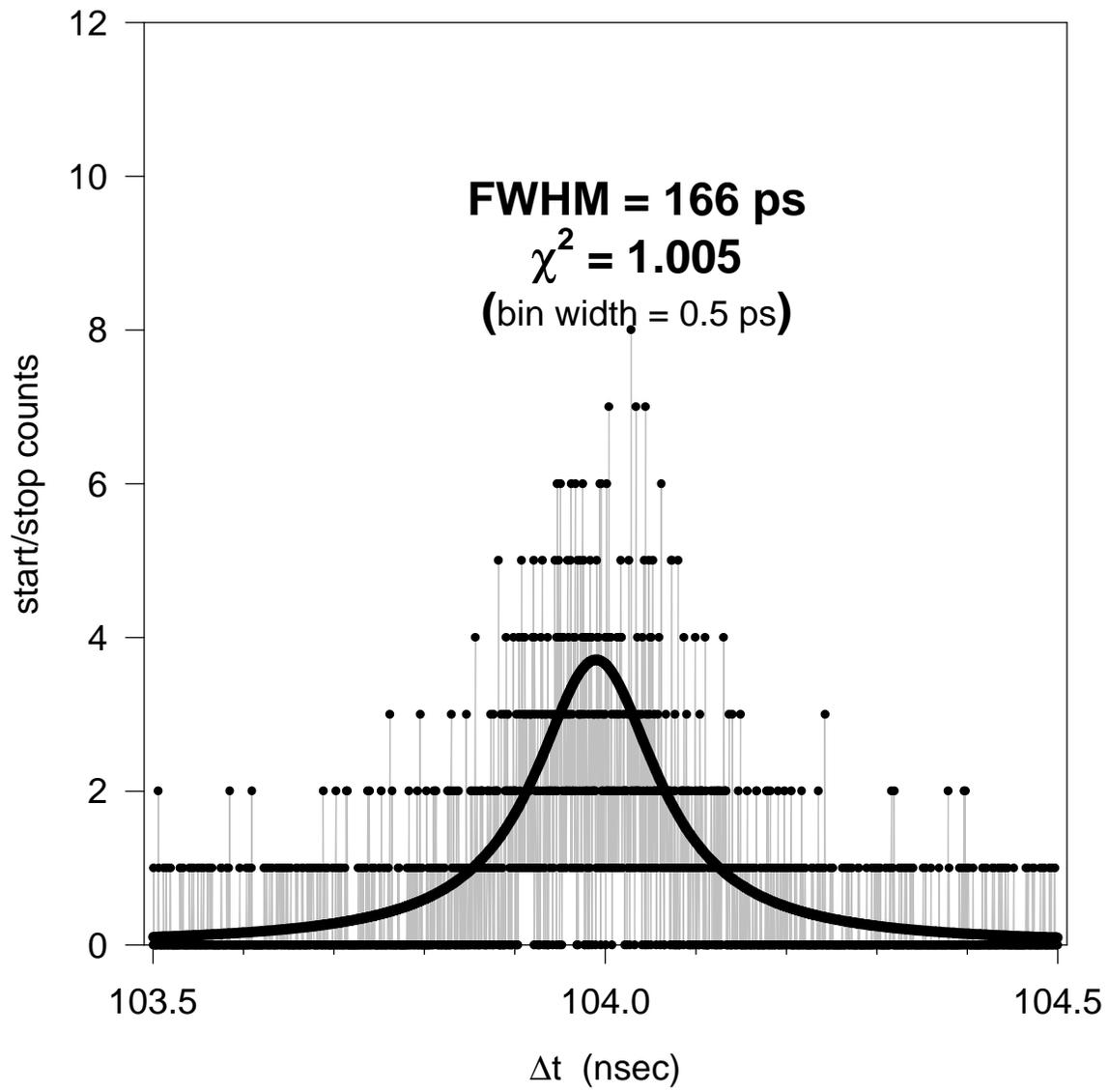

**Figure 12**

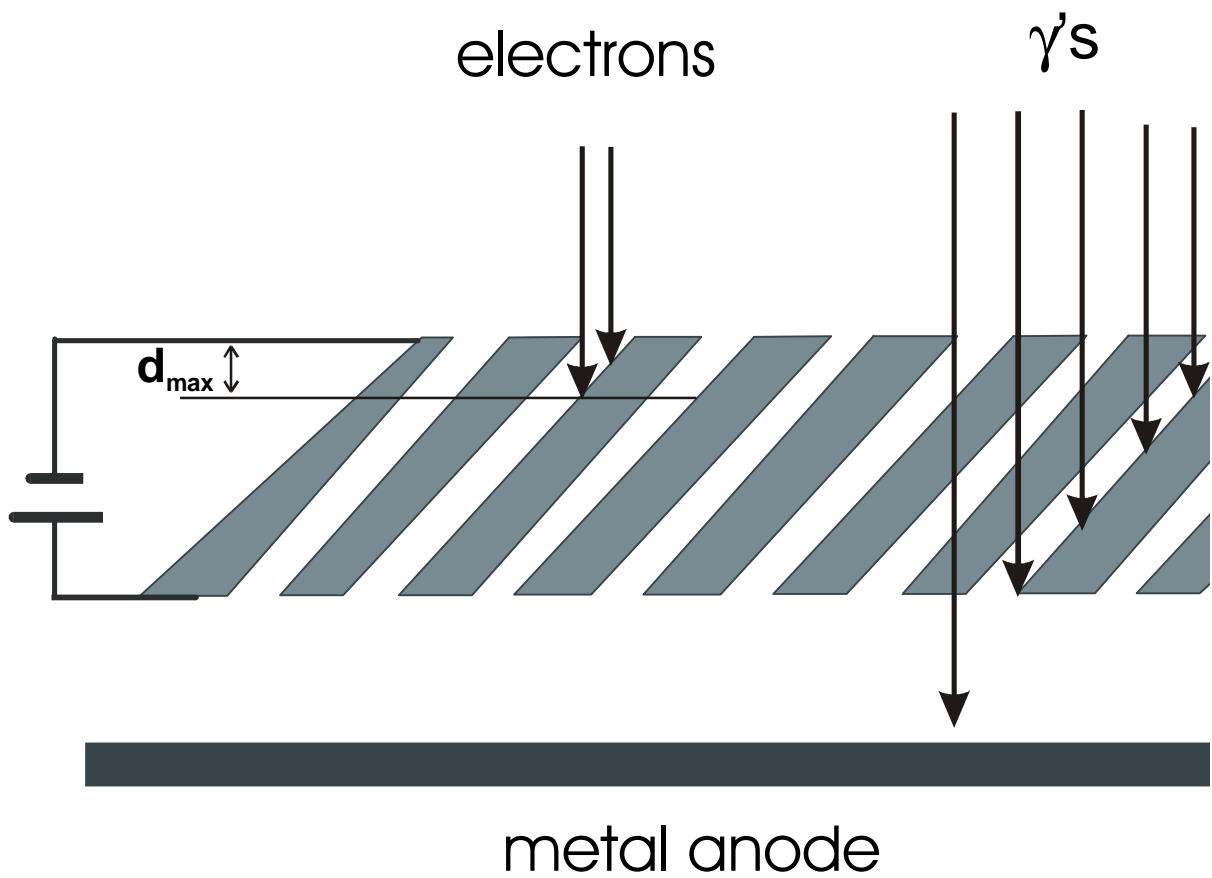

**Figure 13**

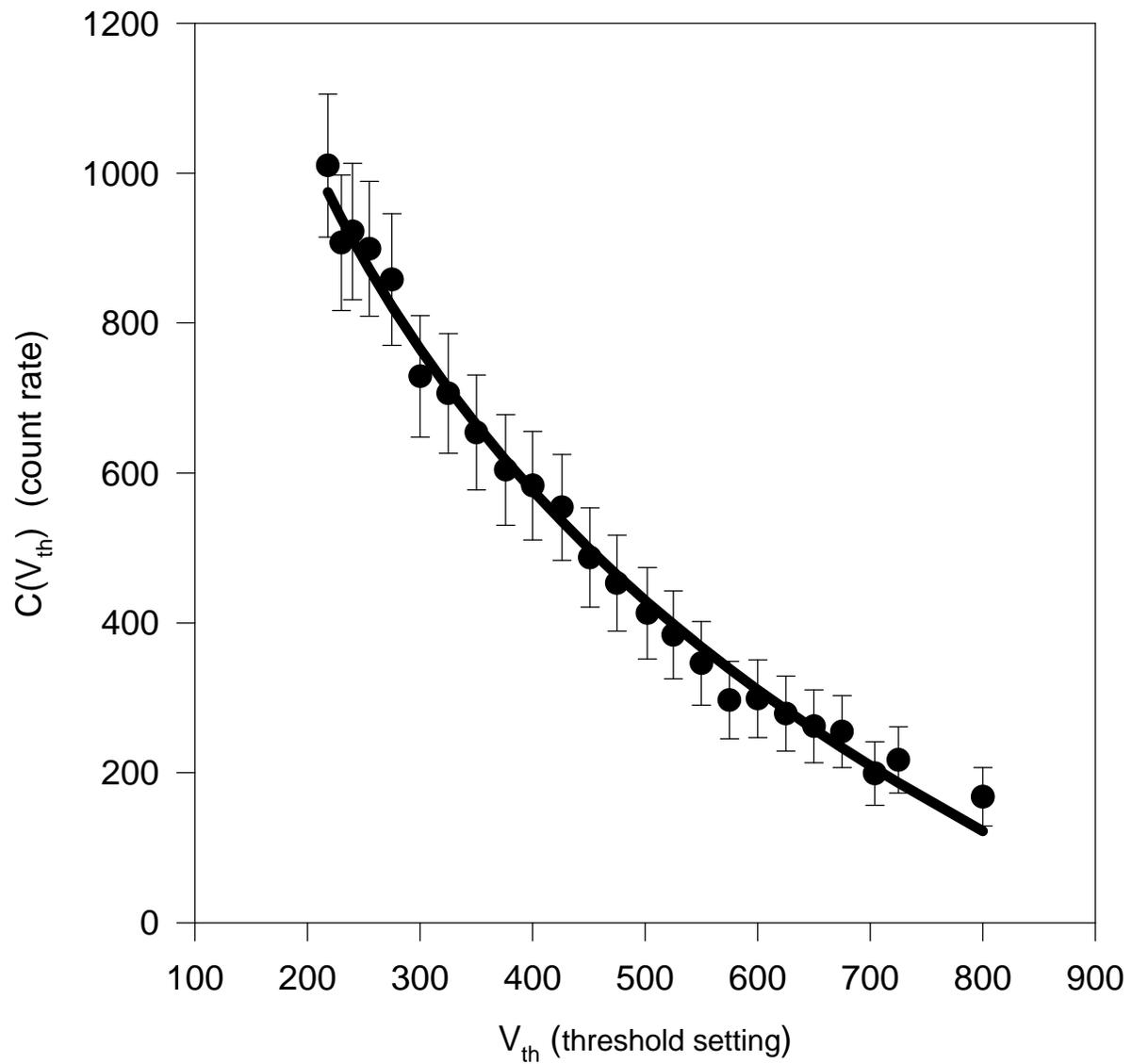

**Figure 14**

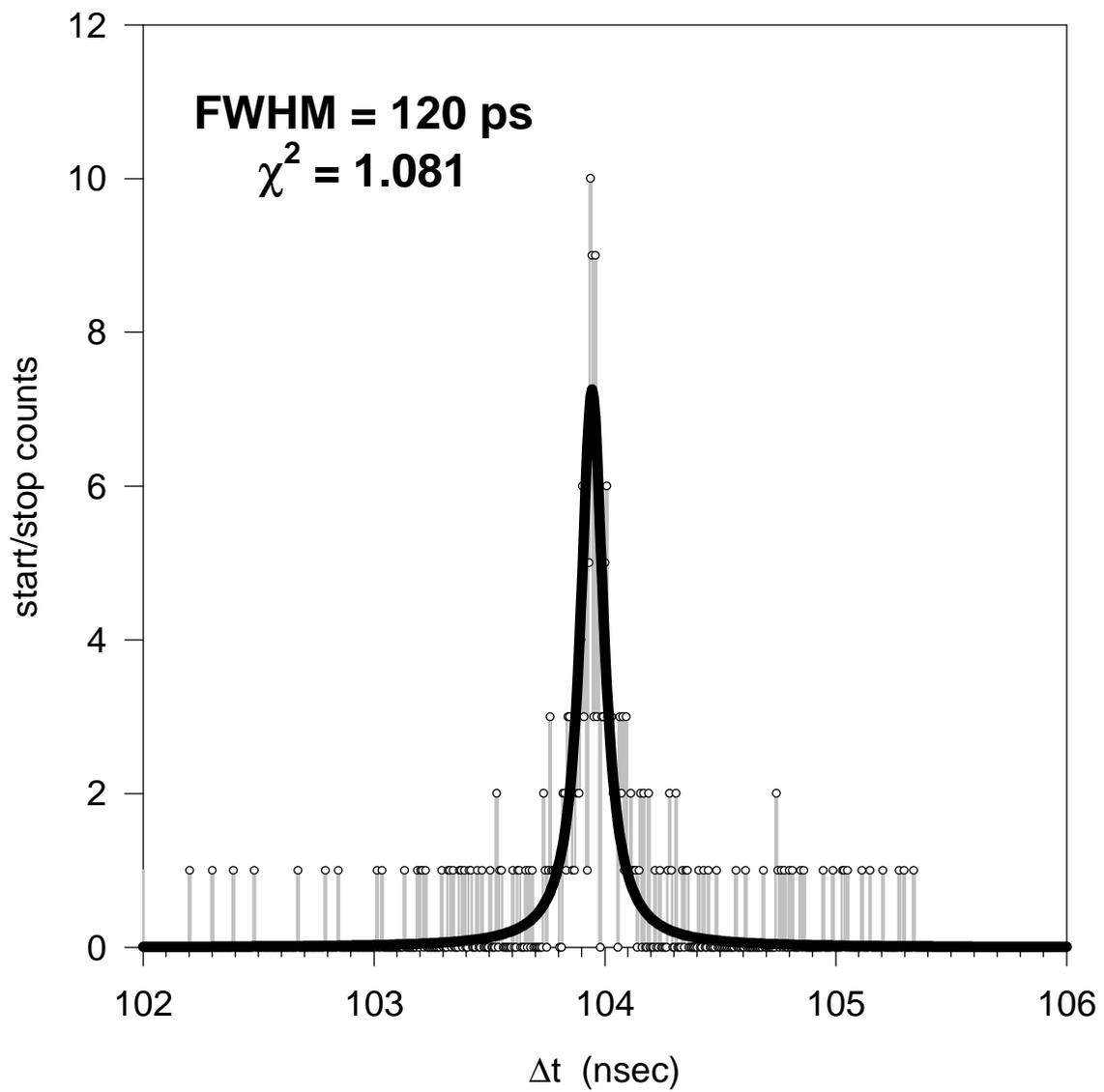

**Figure 15**

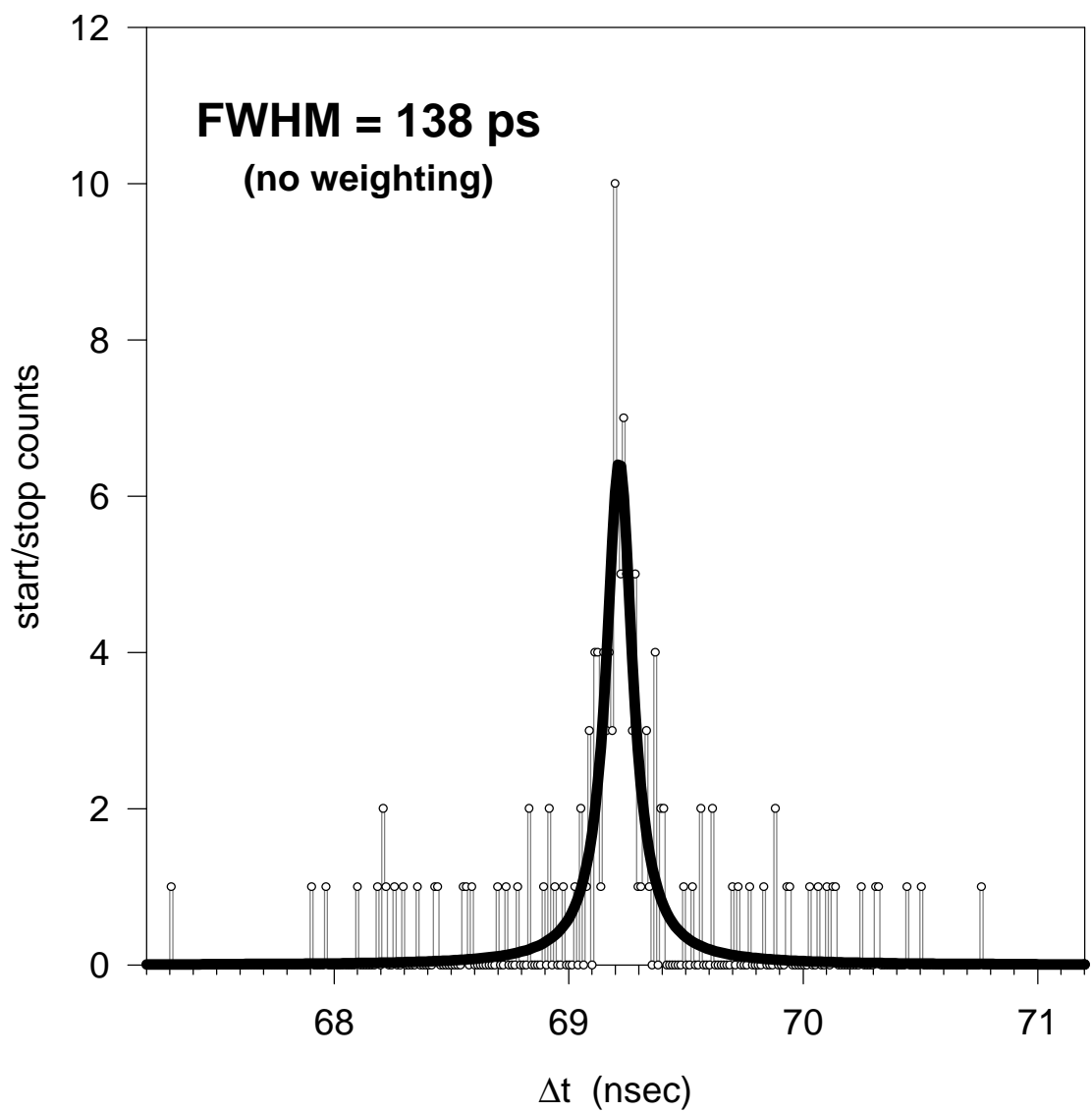

**Figure 16**

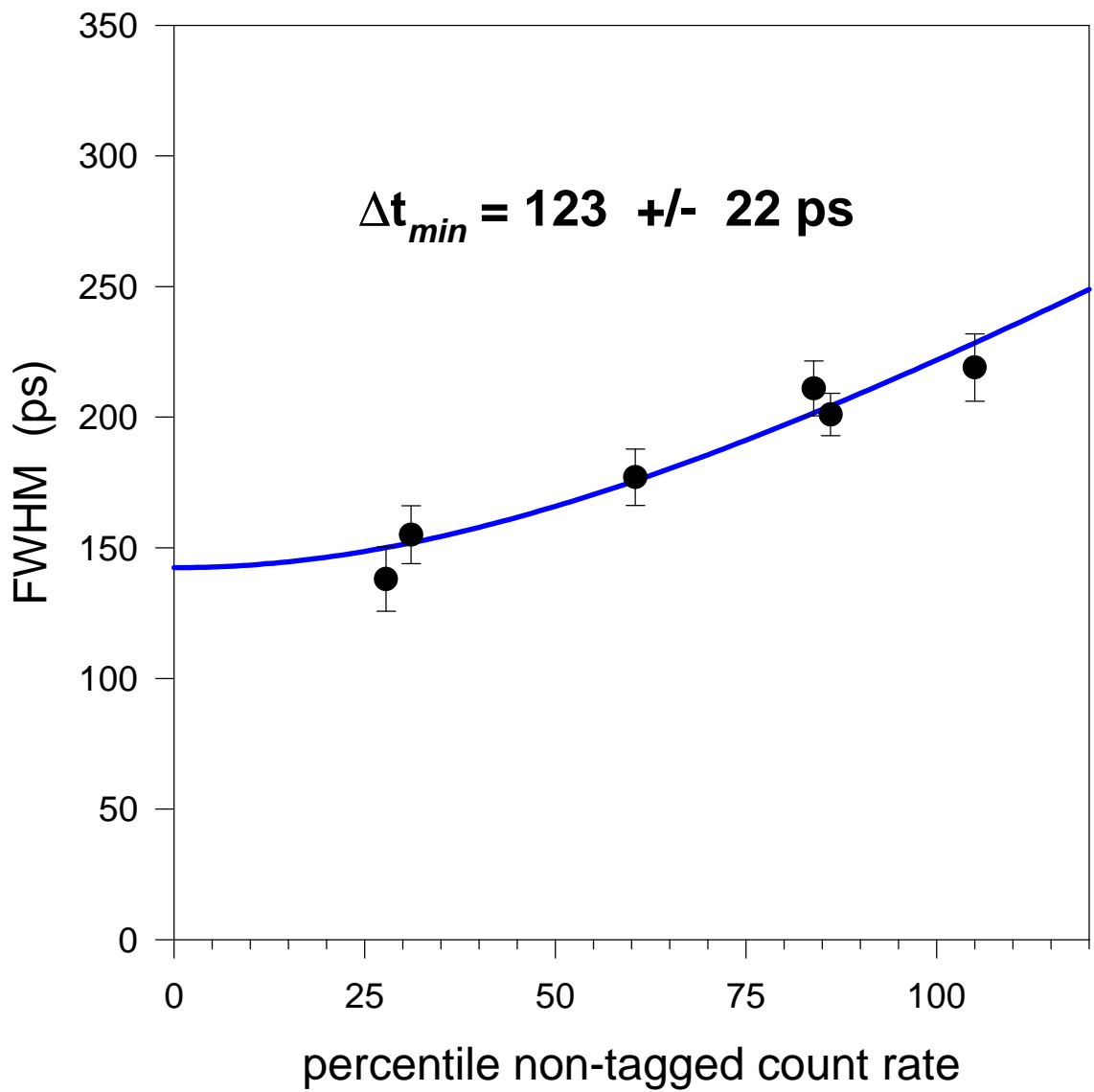

**Figure 17**